# Daytime radiative cooling under extreme weather conditions

Jaesuk Hwang*
Center for Quantum Technologies, National University of Singapore, 3 Science Drive 2, 117543, Singapore

*jhwang@nus.edu.sg

**Abstract**

Radiative cooling, taking advantage of the coldness of the sky, has a potential to be a sustainable alternative to meet cooling needs. The performance of a radiative cooling device is fundamentally limited by the emissivity of the sky, therefore depends heavily on the regional weather conditions. Although the sky emissivity is known to increase with the dew point temperature, this interdependence relates only to clear skies and does not cover extreme weather conditions. Therefore, the feasibility of radiative cooling remains elusive in the equatorial tropical climate. A case study of Singapore is presented where the weather is humid, cloudy and constantly changing. We point out that a high degree of thermal insulation of the radiative cooling system can be effective under such extreme weather conditions. A new method to characterise dynamic sky conditions is presented, namely to measure the sky window emissivity in the zenith direction. We show that a sub-ambient cooling up to 8 °C is possible during daytime and that the cloud base is not a complete blackbody and can be used as a heat sink for radiative cooling.

**Introduction**

The temperature on the earth constantly rises due to the greenhouse effect [1-3]. To cope with the changing climates, the energy expenditure in cooling increases accordingly. The use of compression-based systems, such as air conditioners, accounts for an estimated 20% of electricity consumption in buildings globally [4]. However, the compression-based systems merely move heat from one place to another, causing environmental problems such as the urban heat island (UHI) [5]. Furthermore, a larger consumption of electricity leads to more greenhouse gas emissions, only to accelerate the global warming. Therefore, a zero-energy, passive cooling solution which does not release heat to the environment is needed to slow down the global warming and to support human lives and economy in a sustainable fashion.

To this end, radiative cooling [6-20] is considered a promising solution. When objects on the earth, around 300K temperature, emit towards the extreme coldness of the outer space, at 3K temperature, the object can cool below the ambient temperature passively without any energy input [21, 22]. Radiative cooling devices offer a distinct advantage, namely dissipating heat to the sky, not to the environment, and achieving a sub-ambient temperature, over other passive cooling devices exploiting conduction and convection. Radiative cooling was observed and even used for centuries [23, 24], but only recently it was understood and explained as the longwave imbalance between the thermal radiation of a terrestrial object and the atmospheric radiation [25, 26]. Since the cooling effect was often observed at night, the phenomenon has been often termed the nocturnal cooling [27, 28]. Since the recent demonstration of the daytime radiative cooing under direct sunlight [14, 15], radiative cooling has attracted a broader interest and has been pursued as a sustainable, renewable and eco-friendly alternative energy source.

Such innovative cooling technology would have most impact in hot and humid climates, where a major fraction of energy demand is for cooling. However, radiative cooling in hot and humid climates remains a challenge because radiative cooling depends strongly on the regional climatic conditions, in particular the sky emissivity [10, 26, 29-32]. The sky emissivity depends on the transparency of the "sky window" (from 7.9 to 14 μm, or a smaller range within this interval, from 8 to 13 μm, is often used) [10, 12]. One of the most determining factors for the sky window transparency is known to be the water vapour content in the air of the region, correlated with the local humidity [30, 31]. The sky window in humid climates is therefore more opaque than in dry climates and this makes radiative cooling in an equatorial tropical climate [33-35] or even in tropical and humid sub-tropical climates [36-42] more challenging compared to other parts of the world.

Considering that the sky emissivity increases with the dew point temperature [26], it is understood that the sky emissivity of the equatorial tropical climate would be higher than the other parts of the world. However, the known interpolated curve for the relationship between the sky emissivity and the dew point [26] does not cover the dew point temperature corresponding to equatorial tropical regions. For example, the annual mean relative humidity of Singapore is 83.9 % and the mean temperature is over 26 °C [43] and the corresponding dew point is not part of the available data. Therefore, the feasibility of radiative cooling in the equatorial tropical climate remains elusive.

In addition to the opaque sky window, radiative cooling in the equatorial tropical regions presents further challenges. As illustrated in the left panel of Fig. 1, it rains frequently and there is at all times a significant fraction of cloud coverage. The solar irradiation is intense, with its trajectory passing through the zenith. The weather changes dynamically within a given day. The right panel of Fig. 1 illustrates to contrast the weather condition of a drier region located in a higher altitude, where the sky window is less opaque and the solar irradiation is incident at a more oblique angle. The above mentioned weather conditions near the equator are the common characteristics of the Intertropical Convergence Zone (ITCZ), a narrow belt of convective clouds straddling the equator [44]. In the ITCZ, the trade winds of the northern and southern hemispheres converge, bringing warm and moist air masses. Due to the intense solar irradiation, the air rises to expand and cool, leading to condensation and precipitation from deep convective clouds, accompanying continuous thunderstorms. Although the location changes in time, the ITCZ can be traced as the maximum in time-mean precipitation [45, 46]. In the Southeast Asia, it covers Singapore, Indonesia and southern part of Malaysia. These conditions make the weather of Singapore one of the most challenging for radiative cooling [47].

Such harsh and unpredictable weather conditions render the prediction of the sky emissivity of the equatorial tropical climate even more difficult. The sky emissivity values used for the interpolation curve relate only to clear skies [26] and do not account for the cloudy sky. Although the sky emissivity can be estimated from the atmospheric spectrum [16, 34, 48-50] obtained using the MODTRAN programs [51, 52], the simulation results correspond to the representative values and the variations of the weather at various time scales are inevitably obscured. Furthermore, it was pointed out that downward atmospheric emission may be overestimated in MODTRAN simulations [32]. Therefore, while the sky emissivity obtained using the MODTRAN simulation results may provide a reasonable estimate for a stationary and uniform sky, it is not considered suitable for a dynamic and cloudy sky. Therefore, a method to characterise a dynamic sky is required to understand the radiative cooling performance in the equatorial tropical regions.

There have been considerable efforts to optimise the radiative cooling surface with an ultimate end-goal of a unity reflectivity in the solar spectrum and a unity emissivity in the sky window [53, 54]. However, even with the state-of-the-art radiative cooling substrates, the daytime radiative cooling far below ambient temperature remains challenging. For example, using a highly reflective polymeric coating with 97 % solar reflectivity and 94.2 % infrared emissivity, the sub-ambient radiative cooling of 2 °C below ambient at night and less during daytime was demonstrated in Singapore [35].

To this end, it is pointed out here that a high degree of thermal insulation of the radiative cooling system can be the key in enhancing the performance. The degree of thermal insulation is characterised by the heat transfer coefficient $h_{eff}$, having a linear relation to the parasitic heat gain, $P_{parasitic} = h_{eff}(T_a - T_s)$, where $T_a$ is the ambient air temperature and $T_s$ is the temperature of the radiative cooling substrate. Given the passive nature of radiative cooling, the cooling power occurring from the net outgoing thermal radiation can be lost to any heat gain channels, such as the conduction into the surrounding environment, the convection of air and the radiative exchange with the surroundings and the solar absorption. To suppress these heat gain channels, an infrared-transmitting cover can be used to prevent the convective heat gain [55] and an enclosing structure can be used to reduce the radiative and conductive heat gains from the surrounding to a level of $h_{eff} \sim 10 \ Wm^{-2}K^{-1}$ [18]. However, these measures are not capable of completely suppressing the conductive heat gains, especially into the sky facing side of a radiative cooling surface.

A more effective way to suppress conductive heat gain is to use the aerogels, known to have significantly lower thermal conductivity than traditional insulating materials due to their porous structures [56, 57]. For example, polyethylene aerogel with a thermal conductivity comparable to that of air was used as a convective cover, which transmits thermal infrared radiation. With 18mm thick polyethylene aerogels with a thermal conductivity $k = 28.5 \pm 5 \ Wm^{-1}K^{-1}$, it was demonstrated that the heat transfer coefficient $h_{eff} = 12.9 \ Wm^{-2}K^{-1}$, without the aerogel, was reduced to the heat transfer coefficient $h_{eff} = 1.4 \ Wm^{-2}K^{-1}$, as the aerogel was disposed on a radiative cooling substrate [58].

An avenue to an even stronger suppression of conductive heat gain, beyond the heat insulation level comparable to the equivalent volume of air, is to enclose the radiative cooling substrate within a high vacuum chamber. For example, with a vacuum chamber with a Zinc Selenide window, pumped down to $10^{-5}$ torr, the heat gain was suppressed to a level with the estimated heat transfer coefficient $h_{eff} = 0.2 - 0.3 \ Wm^{-2}K^{-1}$ and the temperature reduction of up to 40 °C was demonstrated at high altitude desert [16]. The vacuum-enhanced radiative cooling system has proven effective for the applications operating at a steady state temperature far below the ambient temperature [59], for example, as a zero-energy refrigerator [16, 60] or an energy harvester [61].

In this work, the feasibility of radiative cooling in the equatorial tropical climate is investigated in detail. To demonstrate the importance of the thermal isolation in the extreme weather conditions, a radiative cooling system is designed and constructed to suppress various heat gain channel, using a high vacuum chamber and a window which transmits less than 2 % of the solar irradiation. The system is field-tested under a cloudy sky of Singapore during the daytime. A new method is presented to characterise the constantly changing sky conditions, namely to directly measure the sky window temperature in the zenith direction. The effective sky emissivity obtained from the measured sky window temperature is compared with the performance of the radiative cooling system.

## Results

### Experimental setup

The experimental apparatus consists of a radiative cooling surface disposed within a vacuum chamber, as shown in Fig 2a. The vacuum chamber is pumped to at least $5 \times 10^{-5}$ mbar to suppress the conductive heat gain to the radiative cooling surface. A Germanium window (Knight Optical) with 10 cm diameter and 1 cm thickness is disposed in the top lid of the vacuum chamber. Germanium was chosen because it is highly absorptive from deep UV to near IR wavelength, up to 1.8 $\mu m$, covering a large range of the solar irradiation, but transmits the thermal radiation throughout the sky window. In contrast to the Zinc Selenide window used in Ref [16], the Germanium window relieves the requirement on the solar reflectivity of the radiative cooling surface and widens the choice of materials for the radiative cooling surface. The Germanium window was anti-reflection coated to have 85% average transmissivity from 7 to 14 $\mu m$ range, overlapping the sky window. Fig. 2b shows the transmission spectrum of the AR-coated Germanium window from 280 nm to 2.5 $\mu m$ as the orange solid curve and the 1.5 AM solar irradiance spectrum for reference. The total transmission of the solar irradiation of the AR-coated Germanium window alone is estimated to be 1.9 % (Supplementary Information 1).

As the radiative cooling surface, the 3M™ Enhanced Specular Reflector film (ESR) was used [62] because it is a mass-produced product with known optical properties and an effective radiative cooling substrate [15, 58]. The reflection spectrum from Refs [63, 64] is shown in Fig. 2b. The specular reflector is inherently reflective throughout the visible and near-IR wavelength range, with 98% average reflectivity from 420 nm to 980 nm at near-normal incidence. The infrared spectrum of the specular reflector is largely selective, with an infrared emissivity of 0.60 and a sky window emissivity of 0.79 (Supplementary Information 1). To supplement the reflection beyond 1.8 $\mu m$ wavelength, aluminium foil tape (3M 425) was attached to the backside of the specular reflector. The reflection spectrum of the aluminium foil tape is shown as the grey dotted curve in Fig. 2b. The blue curve in Fig. 2b represents the residual transmission at normal incidence, estimated from the reflectance of the aluminium tape and the transmittance of the Germanium Window. It is estimated that less than 0.2% of the solar radiation is absorbed by the radiative cooling surface, which gives an effective albedo of the system, 0.998.

The radiative cooling surface, 6.5 cm in diameter, was disposed 0.8 cm below the Germanium window and supported on four strands of bare optical fiber to minimise physical contact with solid surfaces. A K-type thermocouple was attached to centre of the backside of the radiative cooling surface. To minimise the radiative exchange with the surrounding, radiation shields have been placed under and around the radiative cooling surface. 9 layers of horizontal shields, a polished aluminium sheet with 10 cm diameter, were disposed under the radiative cooling surface and 3 layers of vertical surfaces, concentric cylinders made from polished aluminium sheets, were disposed to surround the radiative cooling surfaces and the horizontal shields (Supplementary Information 2).

A concentrator structure, in the form of a truncated cone, was placed outside the vacuum chamber around the Germanium window to enhance radiative cooling. Such concentrator enhances radiative cooling by guiding the thermal radiations from the radiative cooling surface to the zenith direction and by obstructing downwelling incident at steeper angles [16, 18, 30, 65, 66]. The concentrator was made from a polished aluminium sheet with a lower diameter of 10.5 cm and an opening angle of 29 °C from the zenith direction (Supplementary Information 2).

**Theoretical Background**

*Modelling atmospheric emissivity*

The atmospheric spectrum can be modelled based on a simple assumption: outside the sky window, the atmosphere is a blackbody at the ambient temperature and throughout the sky window, a single average value of emissivity is assumed [11, 30, 67]. According to this model, the emissivity of the atmosphere $e_a(\eta, \lambda)$ is written as follows:

$$e_{a,nsw}(\eta, \lambda) = 1 \text{ outside the sky window,}$$
$$e_{a,sw}(\eta) \approx 1 - (1 - e_{avg,sw})^{1/\cos\eta} \text{ within the sky window,} \quad (1)$$

where $e_{avg,sw}$ is the average sky window emissivity in the zenith direction, where $\eta$ is the polar angle from the zenith direction and $\lambda$ is the wavelength. $1/\cos\eta$ dependence originates from the angle-dependent path length through the atmosphere, which relates to the air mass. The atmospheric emission spectrum is given by as $e_a(\eta, \lambda) P(\lambda, T_a)$, where $P(\lambda, T_a)$ is the Planck spectrum at ambient temperature $T_a$. It is noted that the regionality and the changing weather condition is represented by the average sky window emissivity $e_{avg,sw}$. We show later that the average sky window emissivity $e_{avg,sw}$ can be directly measured. The advantage of this simple model is that based only on two measurable parameters, the ambient temperature $T_a$ and the average sky window emissivity $e_{avg,sw}$, the atmospheric spectrum can be characterised in enough detail to evaluate the key performance parameters of radiative cooling such as the cooling power and the minimum achievable temperature.

*Net cooling power*

The net radiative cooling power $P_{rad}(T_a, T_s)$ as a function of the ambient temperature $T_a$ and the radiative cooling substrate temperature $T_s$ is the difference between the emitted power and the received power at the radiative cooling substrate:

$$P_{rad}(T_a, T_s) = P_1(T_s) - P_2(T_a) - P_3(T_a) - P_{parasitic}(T_a, T_s). \tag{2}$$

$P_1$ is the power radiated from the radiative cooling substrate at temperature $T_s$. $P_2$ is the downwelling absorbed by the radiative cooling substrate within the acceptance angle of the concentrator structure. $P_3$ is the "reciprocal rays" term, which represents the downwelling which would not have been incident on the radiative cooling substrate if there were no concentrator structure. This term is crucial not to overestimate the enhancement of the concentrator structure [68]. $P_{parasitic}(T_a, T_s)$ is the parasitic heat gain given by $P_{parasitic} = h_{eff}(T_a - T_s)$. The net cooling power in equation (2) and the steady state temperature, $T_s$ when $P_{rad}(T_a, T_s) = 0$, can be evaluated with the knowledge of three parameters: the ambient temperature $T_a$, the angle-dependent emissivity spectrum of the radiative cooling substrate $e_s(\eta, \lambda)$ and the average sky window emissivity $e_{avg,sw}$ (Supplementary Information 3).

**Measurement results**

*Calibration measurements*

To test and calibrate the radiative cooling device, an indoors experiment was performed. Exposed to a fixed temperature source, the temperature of the radiative cooling surface was monitored at different chamber pressures. An ice pack at 0 °C was used as a heat sink and disposed over the vacuum chamber to cover entire area of the external facing surface of the Germanium window. The concentrator structure was removed in these measurements. Before each measurement, an aluminium plate was inserted between the Germanium window and the ice pack to prevent any radiative exchange between the radiative cooling surface and the ice pack. When the radiative cooling surface was thermalised to the ambient temperature, measurements were initiated by removing the aluminium plate. The temperature reductions from the initial temperature of 28 °C are presented in Fig. 3. At ambient pressure within the chamber, 6.5 °C reduction from the ambient temperature was observed after around 8 minutes. At 8 mbar, prepared with a scroll pump, no improvement from the atmospheric pressure could be observed. Only from $10^{-1}$ mbar level pressure, achieved using a turbo pump and a partially open valve, the enhanced reduction of the temperature was observed. The trend is clear therefrom that the temperature reduction increases with a lower pressure. At $4.5 \times 10^{-5}$ mbar pressure, the temperature reduction of 14 °C from the ambient temperature is observed within 8 minutes. These results demonstrate that by decreasing the thermal conductivity of air, the conductive heat gain can be suppressed, and a larger temperature reduction from the ambient temperature can be obtained.

The reduction of the thermal conductivity of air as a function of pressure is given by $\frac{k}{k_0} = 1/\left(1 + \frac{CT}{pd}\right)$, where $k_0$ is the thermal conductivity at atmospheric pressure in $Wm^{-1}K^{-1}$, $T$ is the air temperature in $K$, $p$ is the pressure in mbar and $d$ is the characteristic length of the system in $m$ and $C$ is a constant $C = 7.6 \times 10^{-7}$ [69]. The characteristic length $d$ is here taken to be the distance between the radiative cooling surface and the inside-facing surface of the Germanium window, 0.8 cm. At $4.5 \times 10^{-5}$ mbar pressure, the thermal conductivity of the evacuated space within the chamber is $4.1 \times 10^{-5}\ Wm^{-1}K^{-1}$. The corresponding heat transfer coefficient, estimated with $k/d$, is $h_{eff} = 5.1 \times 10^{-3}\ Wm^{-2}K^{-1}$ and the corresponding parasitic heat gain $P_{parasitic}$ at 14 °C temperature reduction is less than 0.1 $W/m^2$. This level of heat gain is smaller than 1 % of the measured cooling power (Supplementary Information 3). Therefore, at 4.5 × $10^{-5}$ mbar pressure, the parasitic heat gain is considered negligible.

Although the heat transfer coefficient $h_{eff}$ is negligible at the low pressure, the radiative exchange with the internal surface of the chamber is still present, mainly due to the non-zero emissivity of the radiation shields and the AR coated surface of the Germanium window. These lead to non-negligible radiative warming of the radiative cooling surface. Assuming $h_{eff} \sim 0$, the equation (2) for the net cooling power for the outdoor radiative cooling measurements is modified to:

$$P_{rad}(T_a, T_s) = P_1(T_s) - P_2(T_a) - P_3(T_a) - P_{cal}(T_a). \qquad (2')$$

This calibration contribution, $P_{cal}(T_a)$, is regarded largely independent on the radiative cooling surface temperature $T_s$ or on the temperature reduction $T_a - T_s$. The calibration power $P_{cal}(T_a)$ is therefore assumed to be a constant number which can be estimated based on the time trajectory measurement of temperature such as in Fig. 3 (Supplementary Information 4).

*Clear sky measurements before sunrise*

Fig.4a is a field measurement of the vacuum-enhanced radiative cooling apparatus. The measurements were performed on March 23rd 2023 from 6:22 AM to 7:08 AM in Kent Ridge, Singapore (1°17'49.6"N 103°46'43.9"E) on the roof of a 6-story building which allows a full hemispherical access of the sky. Before the sunrise, the sky was clear without any visible cloud patches. The humidity was measured to be 88% for the duration of the measurement. Before the beginning of the measurement, the vacuum chamber was pumped down to 4.0 × $10^{-5}$ mbar and the window was blocked with an aluminium plate to thermalise the radiative cooling surface to the ambient temperature. The measurements were initiated by removing the aluminium plate opening the view to the sky.

In Fig. 4a, the orange curve labelled as 'vacuum' is the time trajectory of the temperature of the radiative cooling surface and the topmost black curve labelled as 'ambient air' the temperature of the air around the vacuum chamber. The ambient air temperature data was measured with a PT100 sensor (RS Pro) placed in a louver box. The radiative cooling surface was cooled around 8 °C below the ambient temperature within a few minutes and the reduction increased to 9 °C in 20 minutes.

To evaluate the effect of the vacuum insulation on the radiative cooling performance, simultaneous measurements were made on two setups without any vacuum-enhanced thermal isolation. The blue dotted curve labelled 'bare' is the temperature measurement of the radiative cooling surface, the specular reflector on aluminium tape, without any further thermal insulation, exposed to the air and supported on a Styrofoam base. A K-type thermocouple was attached to the aluminium tape side. The red dot dashed line labelled 'reference' is the temperature measurement of the two layers of scotch tape (3M Long Lasting Scotch Tape) attached to an aluminium foil (Reynolds Wrap aluminium foil) used as the radiative cooling substrate [70]. Similar to the 'bare' configuration, a K-type thermocouple was attached to the aluminium foil side and supported on a Styrofoam base. The reference sample, low-cost and easily reproducible, was used to provide a reference point for Singapore to be compared with other regions. For example, with the reference sample, 7 °C reduction was observed at night in Los Angeles, California [70]. For both the 'bare' and 'reference' configurations, around 3 °C temperature reduction from the ambient temperature was recorded. It is mentioned in passing that both the 'bare' and 'reference' surfaces collected dew [71], maintaining a temperature slightly lower than the dew point temperature of 23.5 °C.

The pink-coloured shaded area of Fig. 4a represents the predicted range of the steady state temperature $T_s$ obtained from equations (1) and (2') using the measured sky window emissivity $e_{avg,sw}$, as will be discussed below in Fig. 4b. The range of the predicted value originates from the uncertainty in the calibration power $P_{cal}(T_a)$ (Supplementary Information 4). It is observed that the steady-state temperature of the radiative cooling surface of the vacuum setup falls within the predicted range.

The topmost panel of Fig. 4b shows the measurement data of the sky window temperature $T_{SW}$ in the zenith direction. An infrared thermometer (IR pyrocouple, Calex PC151LT-0mA) was used with an active response in the 8 to 14 $\mu m$ wavelength range and a field of view 15:1. The sky window temperature measured by the infrared thermometer ranged from - 7.8 to - 7.2 °C. Since the infrared thermometer measures the temperature of the atmosphere selectively within the sky window, the measured temperature is directly related to the transparency of the sky window and is lower than the ambient temperature (Supplementary Information 5).

From the measured sky window temperature $T_{SW}$, the sky window emissivity, $e_{avg,sw}$, can be calculated via $e_{avg,sw} = \int_{SW} d\lambda \varepsilon_{IR} P(\lambda, T_{SW}) / \int_{SW} P(\lambda, T_a) d\lambda$, where $\varepsilon_{IR}$ is the fixed emissivity of the infrared thermometer, 0.95. The sky window emissivity $e_{avg,sw}$ calculated with the sky window temperature values in the first panel of Fig. 4b is shown in the second panel of Fig. 4b. The average sky window emissivity $e_{avg,sw}$ ranges from 0.53 to 0.54. In terms of conventionally defined sky emissivity, these values correspond to 0.887 to 0.890, respectively. To put this value in perspective, 0.13 was used for the average sky window emissivity $e_{avg,sw}$ of the sky in Sydney, Australia in Ref [30], which suggests that the sky window of Singapore is at least three times more opaque than that of Australia. In Ref [72], an infrared thermometer was used as an alternative of a Pyrgeometer to measure the total downwelling. Here, an infrared thermometer which is only responsive to the sky window and has a narrow field of view is used to measure the sky window emissivity $e_{avg,sw}$ in the zenith direction. (Supplementary Information 5)

Here we briefly go back to the shaded area of Fig. 4a, corresponding to the theoretical prediction of the minimum achievable temperature. This was obtained by solving $P_{rad}(T_a, T_s) = 0$ of equation (2'), based on the model of the atmospheric spectrum of equation (1), with the measured values of the ambient temperature $T_a$ and the average sky window emissivity in the zenith direction, $e_{avg,sw}$ in the second panel of Fig. 4b. This way, the minimum achievable temperature can be predicted at each time point of the data.

The third panel from the top of Fig. 4b is the measurement of the total downwelling using a Pyrgeometer (Hukseflux SR05-D1A3). The downwelling, the total longwave radiation from the atmosphere towards the earth, is measured to be from 400.2 to 406.5 $W/m^2$. The longwave imbalance, the difference between the downwelling and the blackbody radiation at the ambient temperature, is about 50 $W/m^2$. Although the cooling power higher than this value can be obtained if a concentrator structure is used [30, 65], this amount of longwave imbalance represents the fundamental limit on the available power by radiative cooling for each region and its sky condition [32]. The downwelling measured by the Pyrgeometer $P_{Pyrgeo}$ is directly related to the conventionally defined sky temperature, via $P_{Pyrgeo} = \sigma T_{sky}^4$. The corresponding sky temperature $T_{sky}$, between 16.9 to 18.0 °C, is shown as the brown curve in the bottom panel of Fig. 4b. This shows that the Singapore sky before sunrise is only several degrees colder than the ambient temperature, around 26 °C. The bottom panel of Fig. 4b shows the comparison between the sky temperatures measured by the infrared thermometer and the Pyrgeometer. The sky temperature for the infrared thermometer was obtained by integrating the atmospheric spectrum according to equation (1), using the measured sky window emissivity $e_{avg,sw}$ (Supplementary Information 5). The sky temperature obtained in this fashion is between 16.6 to 17.1, which agrees reasonably with that from the Pyrgeometer measurement.

It is crucial to note that the agreement between the Pyrgeometer measurement and the infrared thermometer measurement is conditional on the sky being uniform. During the measurements presented in Figs. 4a and 4b, no low-lying cloud patch was immediately visible, although a 98% coverage of the high cloud (8 – 15 km height above sea level) was reported [73]. So the sky can be regarded as reasonably uniform throughout the hemisphere. Any irregular cloud pattern, especially the presence of the low-lying cloud patches, would increase the discrepancy between the two measurements. This is because the Pyrgeometer measurement corresponds to the integration over the hemisphere and over the whole infrared spectrum, so the spectral and angular information are obscured, while the infrared thermometer measures in the zenith direction and within the sky window. For example, the Pyrgeometer measurement would not differentiate between a sky with a single, thick cloud patch blocking the zenith direction and a sky with a widely distributed thin clouds, which would lead to very different radiative cooling performances. Also, the definition of the sky temperature $T_{sky}$ assumes that the atmospheric emission follows a Planck spectrum at that temperature and therefore disregards the spectral structure of the atmospheric spectrum, especially the sky window, such as defined in equation (1) (Supplementary Information 5). Therefore, for the radiative cooling devices with a concentrating structure, the infrared thermometer and the measurement of the sky window temperature $T_{SW}$ provide a more accurate evaluation of the relevant sky condition because it is responsive only within the sky window and a small solid angle around the zenith.

Fig. 4c shows the temperature trajectory in the first 10 minutes of the radiative cooling surface of the vacuum-enhanced setup shown in Fig. 4a. The slope of the temperature trajectory is given by $\frac{dT}{dt} = - P_{rad}(T_a, T_s)/C \frac{m}{A}$, where $C$ is the specific heat capacity and $\frac{m}{A}$ is the mass per unit area of the radiative cooling surface. The cooling power at $T_s = T_a$ is therefore proportional to the initial slope of the temperature trajectory. From the slope of the tangent to the curve, shown as the dotted line, the cooling power is estimated to be 38.1 $W/m^2$. Similar measurements were performed during nights and early mornings, and a cooling power between 25.9 and 40.9 $W/m^2$ was measured (Supplementary Information 3).

*daytime cloudy sky measurements*

Fig.5 is the radiative cooling measurement of the vacuum-enhanced setup in the presence of solar irradiation. The measurements were performed on March 22nd 2023 from 8:44 AM to 01:13 PM. At 01:13PM, it rained. The orange curve labelled as 'vacuum' is the time trajectory of the temperature of the radiative cooling surface and the black topmost curve labelled as 'ambient air' the temperature of the air around the vacuum chamber. The ambient air temperature data was measured with a PT100 sensor (RS Pro) placed in a louver box. To eliminate external influences on the louvre box such as heating of the louvre box from the solar irradiation, the louver box was placed in the shade with an open air flow. The louver box was also elevated by around 1 meter to be away from the heat of the concrete floor.

A Pyranometer (Hukseflux IR02-TR) was used to measure the total solar irradiation incident at the site of the measurement, as shown in the second panel of Fig. 5. The altitude of the sun increases with time throughout the measurement period, being the highest, 89 degrees, at 13:13 PM. The correspondingly increasing trend of the solar irradiation is interrupted by the attenuation of moving clouds. The highest solar irradiation is measured to be 915 $W/m^2$ at 12:49 PM.

An infrared thermometer was used to measure average sky window emissivity $e_{avg,sw}$, shown in the bottom panel of Fig. 5, as explained in detail in Figs. 4a and 4b. It shows generally a two-state behaviour, with the lower level around 0.4 to 0.55 and the higher level around 0.7 to 0.72. In terms of the conventionally defined sky emissivity, these values correspond to 0.837 and 0.894, respectively, for the lower level and 0.941 and 0.947, respectively, for the higher level. The higher level coincides with the presence of cloud patches directly above the measurement site. In this case, the infrared thermometer measures the cloud base temperature, which varies depending on the cloud height, thickness and layer compositions [74, 75]. Our measurement confirms the observations that radiative cooling surface warms up when the cloud screens the clear sky [16, 72, 76]. Since the clouds were constantly moving in and out of the field of view of the radiative cooling setup, the cloud base and the clear sky are not differentiated for the emissivity estimation according to equation (1). In other words, the cloud base temperature measured by the infrared thermometer is taken to be the effective sky window temperature $T_{SW}$ and the sky window emissivity $e_{avg,sw}$ is evaluated also for the cloud base.

The predicted range of lowest achievable temperatures is marked by the shaded area in the first panel of Fig. 5. As explained in Figs. 4a and 4b, they are obtained by solving $P_{rad}(T_a, T_s) = 0$ of equation (2') which is based on the sky window emissivity $e_{avg,sw}$ and the range of calibration powers $P_{cal}(T_a)$. It is noted that the predicted range is determined without the presence of the solar radiation and the deviation is attributed to the solar heating of the Germanium window and the corresponding thermal radiation to the radiative cooling surface (Supplementary Information 6).

During the early hours until 10:00 AM, the temperature reduction from the ambient temperature is around 8 °C below the ambient temperature, generally falling within the predicted range. This is because the direct solar illumination at a low incidence angle is blocked effectively by the concentrator structure so the absorbed solar irradiation $P_{solar}$ is deemed negligible. As the altitude of the sun increases, the direct solar illumination is not completely blocked by the concentrator structure and the temperature of the radiative cooling surface deviates from the predicted range.

Often, a significant fraction of the sky was covered with clouds while the sun was fully exposed, as shown in the insert image of the sky at 11:12 AM, where the altitude of the sun is 59 degrees and the solar irradiation is 840 $W/m^2$. From 11:17 AM, when the altitude of the sun is at 61 degrees, the direct solar illumination is incident fully within the acceptance angle of the concentrator structure. Nevertheless, the radiative cooling surface remained around 3 °C below the ambient temperature.

From 11:23 AM to 12:43 AM, thick clouds were present in the zenith direction, as shown in the image of the sky at 12:04 PM. The altitude of the sun was from 63 to 83 degrees during this period. The radiative cooling surface remains below the ambient temperature by 2 to 4 °C and 6 to 15 $W/m^2$ cooling power was measured. This suggests that the cloud base is not a complete blackbody and colder than the ambient temperature and could act as heat sink for radiative cooling. It is also noted that the measured radiative cooling surface temperature lies towards the lower end of or even lower than the predicted range.

During the short periods at 12:49 PM and 01:02 PM, the radiative cooling surface temperature rises slightly above the ambient temperature. As shown in the image of the sky at 12:43 PM, the cloud was widely distributed around the zenith but a hole in the clouds opened up, exposing the sun for a few minutes. The direct solar irradiation was incident near vertically, with 84 and 87 degrees altitudes, respectively. Since the direct solar irradiation was focused by the concentrator structure [65], the effective intensity at the window and the radiative cooling surface is deemed higher than the measured value of 915 $W/m^2$ at 12:49 PM. However, since less than 0.2% of the solar radiation is absorbed by the radiative cooling surface, the direct solar absorption by the radiative cooling surface is not thought to be the main cause of temperature rise. The focused and vertically incident solar irradiation warmed up the Germanium window over the ambient temperature (Supplementary Information 6). Consequently, due to the radiative warming from the Germanium window, the temperature of the radiative cooling surface rose briefly over the ambient temperature. Although the use of a movable solar reflector which tracks the solar trajectory [18] or removing the concentrator structure could have alleviated the heating of the Germanium window, no further reflector structure was used and the concentrator was kept on to evaluate the performance of a stand-alone, stationary device.

The measurements from 11:23 AM to 12:43 AM shows that the clouds attenuate the direct solar irradiation but at the same time prevent the radiative exchange with a clear sky. Considering that the highest point of the sun path is within 25 degrees polar angle from the zenith direction throughout the year [77, 78] and that the cloud base temperature is lower than the ambient temperature, the cloud base is considered as a viable heat sink for radiative cooling, especially around midday.

The temperatures of the reference sample and the bare sample, discussed in Fig. 4, were measured simultaneously but these samples did not stay below the ambient temperature, especially past around 10:00 AM (Supplementary Information 6). The measurements presented in Fig. 5 demonstrate that the daytime radiative cooling is possible in the equatorial tropical climate of Singapore.

**Conclusion and discussion**

In summary, with a commercial polymeric reflector as the radiative cooling surface, a temperature reduction up to 8 °C from the ambient was achieved during daytime under a cloudy Singapore sky. A new method to monitor the dynamic sky conditions was proposed. Throughout the course of the day, the emissivity of the sky was observed to constantly change in time, mainly due to the dynamics of cloud coverage. The emissivity of the cloud base was measured to be around 0.94 and a temperature reduction up to 3 °C was observed with the cloud base as a heat sink for radiative cooling. Assuming a complete suppression of the heat gain channels, 62 $W/m^2$ cooling power and 19 °C temperature reduction are predicted in Singapore using the same radiative cooling surface used in this work (Supplementary Information 7). Based on these findings, it is concluded that daytime radiative cooling in the equatorial tropical climate is feasible and that the main parameter for optimisation is the suppression of the heat gain channels. Considering that Singapore is considered one of the most adverse weathers for radiative cooling [47], our results speak for the universal applicability of the technique.

**Acknowledgement**

This research is supported by the National Research Foundation, Singapore and A*STAR under its CQT Bridging Grant. The author thanks Prof. Christian Kurtsiefer for hosting him at the Centre for Quantum Technologies at NUS. The author thanks Dr. Tan Peng Kian for the help at the NUS observatory, Dr. Poh Hou Shun for the loan of the weather station and Mr. Lian Chorng Wang for technical support. The author thanks Dr. Kwan Bum Choi and Ms. Elisaveta Ungur at the SERIS (Solar Energy Research Institute of Singapore) of NUS for the help with FT-IR and UV-Vis measurements. The author thanks Dr. Yu-Hung Lien for fruitful discussions.

**References**

[1] P. Nema, S. Nema, P. Roy, An overview of global climate changing in current scenario and mitigation action, Renewable & Sustainable Energy Reviews, 16 (2012) 2329-2336.
[2] Y. Xu, V. Ramanathan, D.G. Victor, Global warming will happen faster than we think, Nature Publishing Group UK London, 2018.
[3] I.P.o.C. Change, Global warming of 1.5° C, World Meteorological Organization: Geneva, Switzerland, DOI (2018).
[4] I.E. Agency, The Future of Cooling: Opportunities for energy- efficient air conditioning, 2018.
[5] H.E. Landsberg, The urban climate, Academic press1981.
[6] F. Trombe, Perspectives sur l'utilisation des rayonnements solaires et terrestres dans certaines régions du monde. , Rev. Gén. Therm. , 6 (1967) 1285–1314

[7] A.W. Harrison, M.R. Walton, RADIATIVE COOLING OF TIO2 WHITE PAINT, Solar Energy, 20 (1978) 185-188.
[8] C.G. Granqvist, RADIATIVE HEATING AND COOLING WITH SPECTRALLY SELECTIVE SURFACES, Applied Optics, 20 (1981) 2606-2615.
[9] A. Hjortsberg, C.G. Granqvist, RADIATIVE COOLING WITH SELECTIVELY EMITTING ETHYLENE GAS, Applied Physics Letters, 39 (1981) 507-509.
[10] T.S. Eriksson, C.G. Granqvist, RADIATIVE COOLING COMPUTED FOR MODEL ATMOSPHERES, Applied Optics, 21 (1982) 4381-4388.
[11] E.M. Lushiku, A. Hjortsberg, C.G. Granqvist, RADIATIVE COOLING WITH SELECTIVELY INFRARED-EMITTING AMMONIA GAS, Journal of Applied Physics, 53 (1982) 5526-5530.
[12] P. Berdahl, M. Martin, F. Sakkal, THERMAL PERFORMANCE OF RADIATIVE COOLING PANELS, International Journal of Heat and Mass Transfer, 26 (1983) 871-880.
[13] A.R. Gentle, G.B. Smith, Radiative Heat Pumping from the Earth Using Surface Phonon Resonant Nanoparticles, Nano Letters, 10 (2010) 373-379.
[14] A.P. Raman, M. Abou Anoma, L.X. Zhu, E. Rephaeli, S.H. Fan, Passive radiative cooling below ambient air temperature under direct sunlight, Nature, 515 (2014) 540-+.
[15] A.R. Gentle, G.B. Smith, A Subambient Open Roof Surface under the Mid-Summer Sun, Advanced Science, 2 (2015).
[16] Z. Chen, L.X. Zhu, A. Raman, S.H. Fan, Radiative cooling to deep sub-freezing temperatures through a 24-h day-night cycle, Nature Communications, 7 (2016).
[17] J.L. Kou, Z. Jurado, Z. Chen, S.H. Fan, A.J. Minnich, Daytime Radiative Cooling Using Near-Black Infrared Emitters, Acs Photonics, 4 (2017) 626-630.
[18] B. Bhatia, A. Leroy, Y.C. Shen, L. Zhao, M. Gianello, D.H. Li, T. Gu, J.J. Hu, M. Soljacic, E.N. Wang, Passive directional sub-ambient daytime radiative cooling, Nature Communications, 9 (2018).
[19] D.L. Zhao, A. Aili, Y. Zhai, J.T. Lu, D. Kidd, G. Tan, X.B. Yin, R.G. Yang, Subambient Cooling of Water: Toward Real-World Applications of Daytime Radiative Cooling, Joule, 3 (2019) 111-123.
[20] L. Zhou, H. Song, J. Liang, M. Singer, M. Zhou, E. Stegenburgs, N. Zhang, C. Xu, T. Ng, Z. Yu, A polydimethylsiloxane-coated metal structure for all-day radiative cooling, Nature Sustainability, 2 (2019) 718-724.
[21] W. Li, S. Fan, Radiative cooling: harvesting the coldness of the universe, Optics and Photonics News, 30 (2019) 32-39.
[22] M.M. Hossain, M. Gu, Radiative cooling: principles, progress, and potentials, Advanced Science, 3 (2016) 1500360.
[23] The Persian ice house, or how to make ice in the desert https://www.fieldstudyoftheworld.com/persian-ice-house-how-make-ice-desert/.
[24] R. Barker, XXII. The process of making ice in the East Indies. By Sir Robert Barker, FRS in a letter to Dr. Brocklesby, Philosophical Transactions of the Royal Society of London, DOI (1775) 252-257.
[25] D. Brunt, Notes on radiation in the atmosphere. I, Quarterly Journal of the Royal Meteorological Society, 58 (1932) 389-420.
[26] P. Berdahl, Retrospective on the resource for radiative cooling, Journal of Photonics for Energy, 11 (2021) 042106-042106.
[27] M.I. Ahmad, H. Jarimi, S. Riffat, Nocturnal cooling technology for building applications, Springer2019.

[28] K. Nwaigwe, C. Okoronkwo, N.V. Ogueke, E. Anyanwu, Review of nocturnal cooling systems, International Journal of Energy for a Clean Environment, 11 (2010).
[29] G.B. Smith, Green Nanotechnology,  Conference on Nanostructured Thin Films IV, San Diego, CA, 2011.
[30] G.B. Smith, Amplified radiative cooling via optimised combinations of aperture geometry and spectral emittance profiles of surfaces and the atmosphere, Solar Energy Materials and Solar Cells, 93 (2009) 1696-1701.
[31] J. Feng, K. Gao, M. Santamouris, K.W. Shah, G. Ranzi, Dynamic impact of climate on the performance of daytime radiative cooling materials, Solar Energy Materials and Solar Cells, 208 (2020).
[32] J. Mandal, X. Huang, A.P. Raman, Accurately Quantifying Clear-Sky Radiative Cooling Potentials: A Temperature Correction to the Transmittance-Based Approximation, Atmosphere, 12 (2021) 1195.
[33] D. Han, J.P. Fei, H. Li, B.F. Ng, The criteria to achieving sub-ambient radiative cooling and its limits in tropical daytime, Building and Environment, 221 (2022).
[34] D. Han, B.F. Ng, M.P. Wan, Preliminary study of passive radiative cooling under Singapore's tropical climate, Solar Energy Materials and Solar Cells, 206 (2020).
[35] D. Han, J. Fei, J. Mandal, Z. Liu, H. Li, A.P. Raman, B.F. Ng, Sub-ambient radiative cooling under tropical climate using highly reflective polymeric coating, Solar Energy Materials and Solar Cells, 240 (2022) 111723.
[36] S.Y. Jeong, C.Y. Tso, J. Ha, Y.M. Wong, C.Y.H. Chao, B.L. Huang, H.H. Qiu, Field investigation of a photonic multi-layered TiO2 passive radiative cooler in sub-tropical climate, Renewable Energy, 146 (2020) 44-55.
[37] S.Y. Jeong, C.Y. Tso, Y.M. Wong, C.Y.H. Chao, B. Huang, Daytime passive radiative cooling by ultra emissive bio-inspired polymeric surface, Solar Energy Materials and Solar Cells, 206 (2020).
[38] C.Y. Tso, K.C. Chan, C.Y. Chao, A field investigation of passive radiative cooling under Hong Kong's climate, Renewable energy, 106 (2017) 52-61.
[39] R.Y. Wong, C. Tso, S. Jeong, S. Fu, C.Y. Chao, Critical sky temperatures for passive radiative cooling, Renewable Energy, 211 (2023) 214-226.
[40] H. Zhong, P. Zhang, Y. Li, X. Yang, Y. Zhao, Z. Wang, Highly solar-reflective structures for daytime radiative cooling under high humidity, ACS Applied Materials & Interfaces, 12 (2020) 51409-51417.
[41] J. Liu, Z. Zhou, D. Zhang, S. Jiao, Y. Zhang, L. Luo, Z. Zhang, F. Gao, Field investigation and performance evaluation of sub-ambient radiative cooling in low latitude seaside, Renewable Energy, 155 (2020) 90-99.
[42] M. Dong, N. Chen, X. Zhao, S. Fan, Z. Chen, Nighttime radiative cooling in hot and humid climates, Optics express, 27 (2019) 31587-31598.
[43] Climate of Singapore, Metrological Service Singapore http://www.weather.gov.sg/climate-climate-of-singapore/.
[44] T. Schneider, T. Bischoff, G.H. Haug, Migrations and dynamics of the intertropical convergence zone, Nature, 513 (2014) 45-53.
[45] D.J. Seidel, Q. Fu, W.J. Randel, T.J. Reichler, Widening of the tropical belt in a changing climate, Nature geoscience, 1 (2008) 21-24.
[46] G. Yancheva, N.R. Nowaczyk, J. Mingram, P. Dulski, G. Schettler, J.F. Negendank, J. Liu, D.M. Sigman, L.C. Peterson, G.H. Haug, Influence of the intertropical convergence zone on the East Asian monsoon, Nature, 445 (2007) 74-77.

[47] L. Carlosena, Á. Ruiz-Pardo, E.Á. Rodríguez-Jara, M. Santamouris, Worldwide potential of emissive materials based radiative cooling technologies to mitigate urban overheating, Building and Environment, DOI (2023) 110694.
[48] C. Liu, Y. Wu, B. Wang, C. Zhao, H. Bao, Effect of atmospheric water vapor on radiative cooling performance of different surfaces, Solar Energy, 183 (2019) 218-225.
[49] X. Yu, C. Chen, A simulation study for comparing the cooling performance of different daytime radiative cooling materials, Solar Energy Materials and Solar Cells, 209 (2020) 110459.
[50] J. Feng, M. Santamouris, K.W. Shah, G. Ranzi, Thermal analysis in daytime radiative cooling,  IOP Conference Series: Materials Science and Engineering, IOP Publishing, 2019, pp. 072064.
[51] D. Archer, MODTRAN, http://climatemodels.uchicago.edu/modtran/, University of Chicago.
[52] A. Berk, P. Conforti, R. Kennett, T. Perkins, F. Hawes, J. van den Bosch, MODTRAN (R) 6: A major upgrade of the MODTRAN (R) radiative transfer code,  20th SPIE Conference on Algorithms and Technologies for Multispectral, Hyperspectral, and Ultraspectral Imagery, Baltimore, MD, 2014.
[53] X.X. Yu, J.Q. Chan, C. Chen, Review of radiative cooling materials: Performance evaluation and design approaches, Nano Energy, 88 (2021).
[54] M.J. Chen, D. Pang, X.Y. Chen, H.J. Yan, Y. Yang, Passive daytime radiative cooling: Fundamentals, material designs, and applications, Ecomat, 4 (2022).
[55] J. Zhang, J.J. Yuan, J.W. Liu, Z.H. Zhou, J.Y. Sui, J.C. Xing, J. Zuo, Cover shields for sub-ambient radiative cooling: A literature review, Renewable & Sustainable Energy Reviews, 143 (2021).
[56] A.S. Dorcheh, M.H. Abbasi, Silica aerogel; synthesis, properties and characterization, Journal of Materials Processing Technology, 199 (2008) 10-26.
[57] M. Schmidt, F. Schwertfeger, Applications for silica aerogel products, Journal of Non-Crystalline Solids, 225 (1998) 364-368.
[58] A. Leroy, B. Bhatia, C.C. Kelsall, A. Castillejo-Cuberos, M. Di Capua H, L. Zhao, L. Zhang, A. Guzman, E. Wang, High-performance subambient radiative cooling enabled by optically selective and thermally insulating polyethylene aerogel, Science advances, 5 (2019) eaat9480.
[59] M. Hu, B. Zhao, J. Cao, Q. Wang, S. Riffat, Y. Su, G. Pei, Effect of vacuum scheme on radiative sky cooling performance, Applied Thermal Engineering, 219 (2023) 119657.
[60] C. Granqvist, A. Hjortsberg, Surfaces for radiative cooling: Silicon monoxide films on aluminum, Applied Physics Letters, 36 (1980) 139-141.
[61] X. Ao, B. Li, B. Zhao, M. Hu, H. Ren, H. Yang, J. Liu, J. Cao, J. Feng, Y. Yang, Self-adaptive integration of photothermal and radiative cooling for continuous energy harvesting from the sun and outer space, Proceedings of the National Academy of Sciences, 119 (2022) e2120557119.
[62] M.F. Weber, C.A. Stover, L.R. Gilbert, T.J. Nevitt, A.J. Ouderkirk, Giant birefringent optics in multilayer polymer mirrors, Science, 287 (2000) 2451-2456.
[63] Perkin Elmer Field Application Report, July 2023 https://resources.perkinelmer.com/lab-solutions/resources/docs/far_measurement-of-enhanced-specular-reflector-films-using-lambda-1050-and-ura-accessory-012190_01.pdf.
[64] M. Janecek, Reflectivity spectra for commonly used reflectors, IEEE Transactions on Nuclear Science, 59 (2012) 490-497.

[65] A. Gentle, A. Nuhoglu, M. Arnold, G. Smith, 3D printable optical structures for sub-ambient sky cooling, Thermal Radiation Management for Energy Applications, SPIE, 2017, pp. 16-24.
[66] S. Catalanotti, V. Cuomo, G. Piro, D. Ruggi, V. Silvestrini, G. Troise, The radiative cooling of selective surfaces, Solar Energy, 17 (1975) 83-89.
[67] M. Martin, P. Berdahl, Summary of results from the spectral and angular sky radiation measurement program, Solar Energy, 33 (1984) 241-252.
[68] M. Dong, L. Zhu, B. Jiang, S. Fan, Z. Chen, Concentrated radiative cooling and its constraint from reciprocity, Optics Express, 30 (2022) 275-285.
[69] Fluid Flow Databook Section 410.2, Genium Publishing, General Electric, May 1982.
[70] X. Huang, J. Mandal, A.P. Raman, Do-it-yourself radiative cooler as a radiative cooling standard and cooling component for device design, Journal of Photonics for Energy, 12 (2022).
[71] E. Simsek, J. Mandal, A.P. Raman, L. Pilon, Dropwise condensation reduces selectivity of sky-facing radiative cooling surfaces, International Journal of Heat and Mass Transfer, 198 (2022) 123399.
[72] J.L.C. Aguilar, A.R. Gentle, G.B. Smith, D. Chen, A method to measure total atmospheric long-wave down-welling radiation using a low cost infrared thermometer tilted to the vertical, Energy, 81 (2015) 233-244.
[73] https://www.meteoblue.com/.
[74] A. Maghrabi, R. Clay, D. Riordan, Detecting cloud with a simple infra-red sensor, Transactions of the Royal Society of South Australia, 133 (2009) 164-171.
[75] D. Riordan, R. Clay, A. Maghrabi, B. Dawson, R. Pace, N. Wild, Cloud base temperature measurements using a simple longwave infrared cloud detection system, Journal of Geophysical Research: Atmospheres, 110 (2005).
[76] A. Gentle, G. Smith, Performance comparisons of sky window spectral selective and high emittance radiant cooling systems under varying atmospheric conditions, Proceedings Solar2010, The 48th AuSES Annual Conference, 2010.
[77] Singapore Sun Path Diagram, https://www.gaisma.com/en/location/singapore.html.
[78] Sun Direction in Singapore, https://sun-direction.com/city/54050,singapore/, DOI.

**Figure 1.** Left: illustration of the weather condition near the equator relevant for radiative cooling. While the solar illumination is intense, the outgoing thermal radiation is highly attenuated by the opaque atmosphere and the heavy clouds. Right: illustration of the weather condition at higher altitude to contrast with that near the equator. The atmosphere is less opaque and the solar illumination, incident at a more oblique angle.

**Figure 2.** (a) The experimental setup. A radiative cooling surface, a 3M specular reflector on aluminium tape is enclosed within a vacuum chamber pumped to at least $5 \times 10^{-5}$ mbar to suppress the conductive heat gain. An AR-coated Germanium window is used to transmit thermal radiation to and from the radiative cooling surface. A concentrator structure in a truncated cone shape is used to amplify the radiative cooling effect. (b) The spectrum in the UV, visible and near-IR wavelength ranges of the Germanium window (orange dotted line) and the total transmissivity considering the Germanium window and the radiative cooling surface (blue line).

The effective albedo of the system is estimated to be 0.981. 1.5 AM solar irradiance spectrum is also shown for comparison (gray line).

**Figure 3.** Radiative cooling measurements with different chamber pressures with an ice pack disposed over the vacuum chamber as the radiative heat sink. The minimum achievable temperature of the radiative cooling surface decreases as the pressure decreases.

**Figure 4.** (a) Radiative cooling measurements under a clear sky before sunrise. The 'vacuum' (orange line) is the temperature of the radiative cooling surface and the 'ambient air' (black line) is the temperature of the air around the vacuum chamber. 9 °C temperature reduction from the ambient is observed. This performance is compared with samples exposed to the air without any further thermal insulation. The 'bare' sample (blue dotted line) is the radiative cooling surface, 3M specular reflector on aluminium tape and the 'reference' sample (red dot-dashed line) includes the two layers of scotch tape attached to an aluminium foil. The shaded area represents the predicted range of the steady state temperature. (b) The topmost panel shows the measurement data of the sky window temperature in the zenith direction, measured with an infrared thermometer. The second panel is the sky window emissivity in the zenith direction calculated from the measured sky window temperature. The third panel from the top is the measurement of the total downwelling using a Pyrgeometer. The bottom panel shows the sky temperature estimated using the infrared thermometer measurements and the sky temperature obtained by the Pyrgeometer in comparison. (c) The temperature reduction from the ambient temperature in the first 10 minutes of the measurement in Fig. 3(a). From the slope of the tangent to the curve, shown as the dotted line, the cooling power is estimated to be 38.1 $W/m^2$.

**Figure 5.** The top panel shows the radiative cooling measurements under a cloudy sky during daytime. The 'vacuum' (orange line) is the temperature of the radiative cooling surface and the 'ambient air' (black line) is the temperature of the air around the vacuum chamber. The insets are the photos of the sky above the measurement site, showing the distribution of the clouds. The middle panel is the total solar irradiation incident at the site of the measurement measured with a Pyranometer. The increasing trend of the solar irradiation is often disrupted when blocked by clouds. The bottom panel shows the average sky window emissivity measured with an infrared thermometer, which shows a general two-level behaviour. The high level of sky window emissivity corresponds to the presence of clouds above the measurement site. Based on the measured sky window emissivity, the predicted range of the steady state temperature is evaluated in the top panel as the shaded area. See the text for more details.

**Fig. 1**

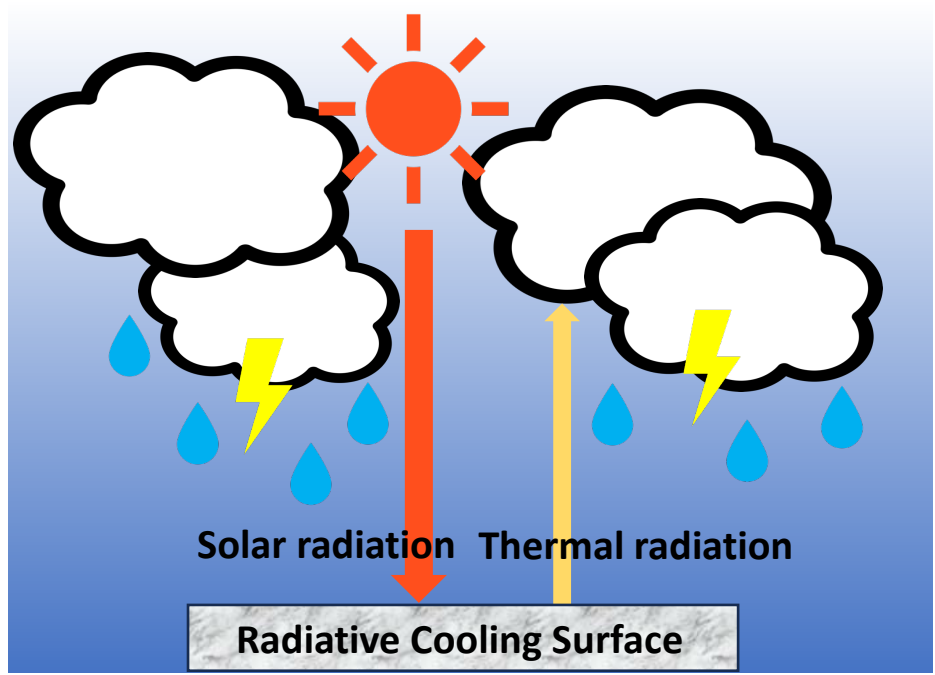 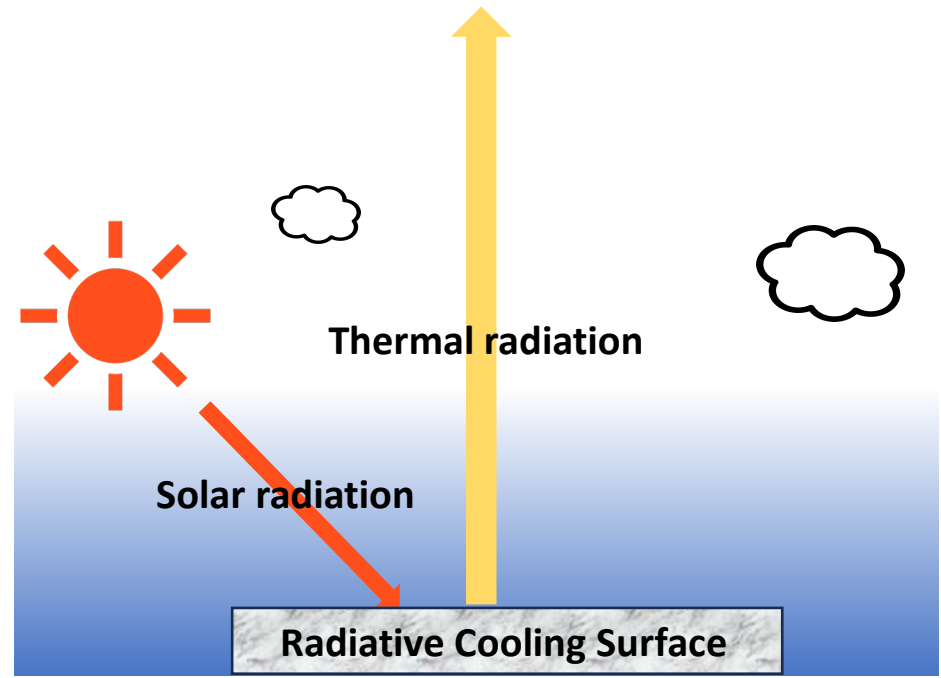

**Fig. 2a**

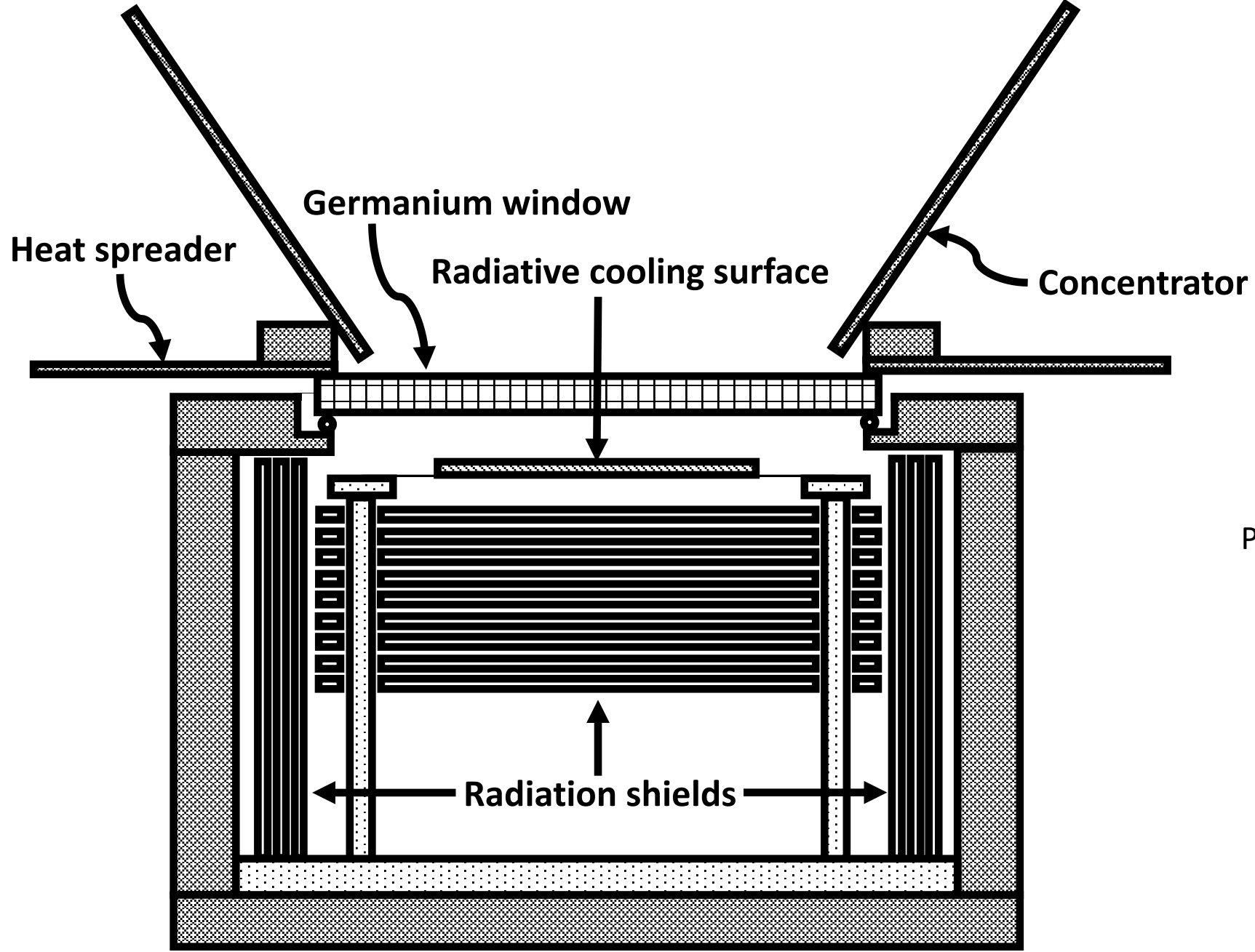

Photo?

**Fig. 2b**

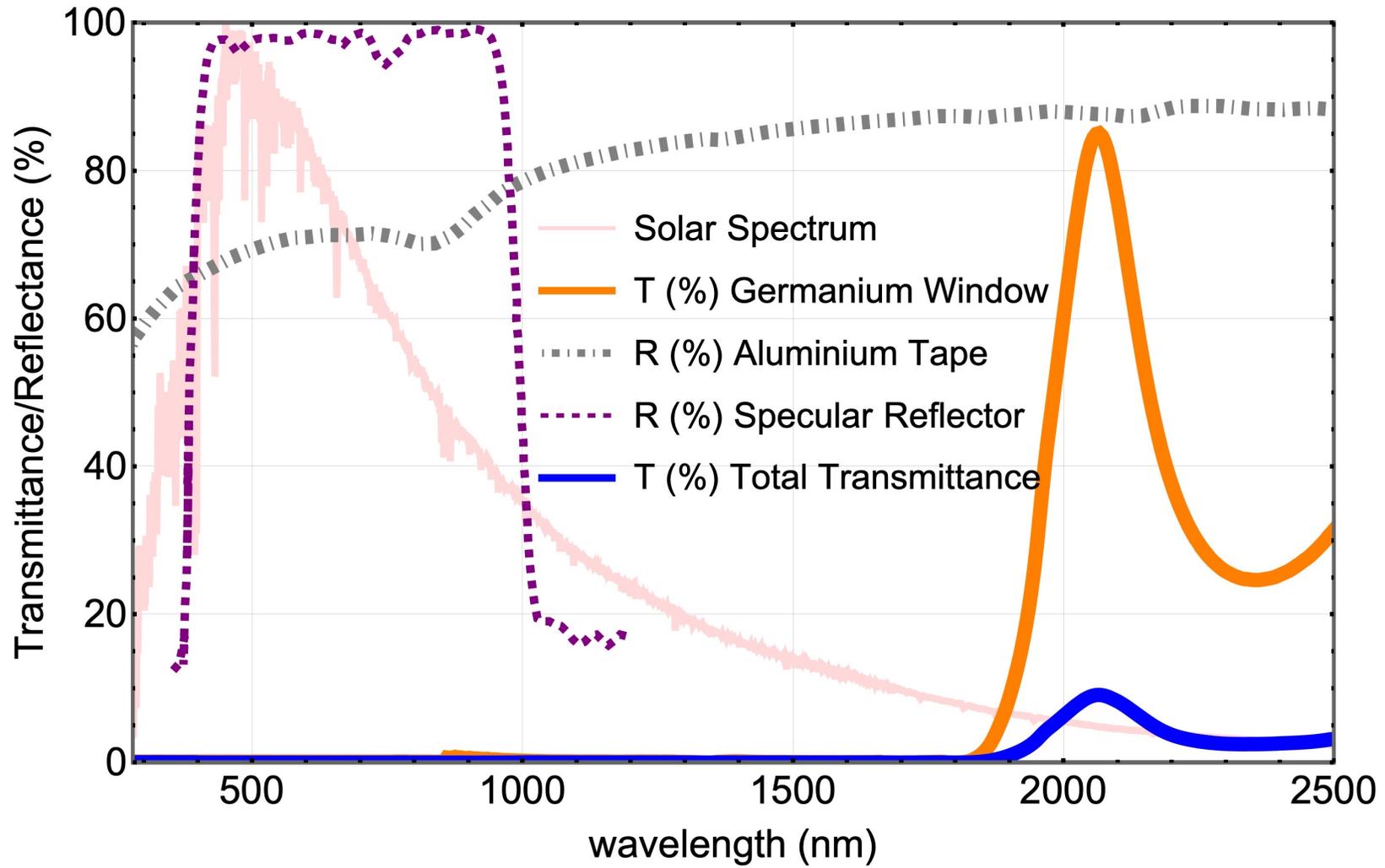

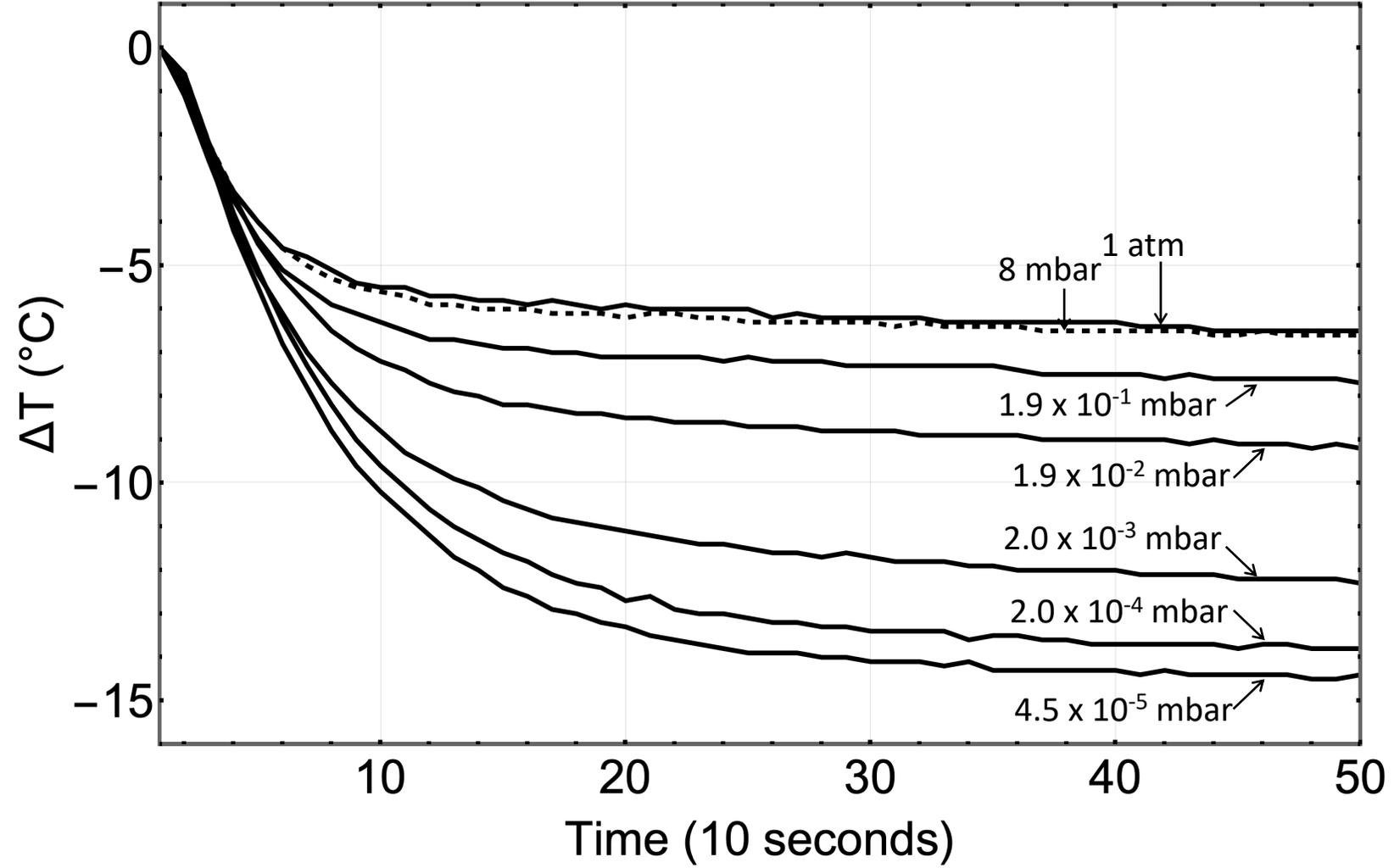

Fig. 3

**Fig. 4a**

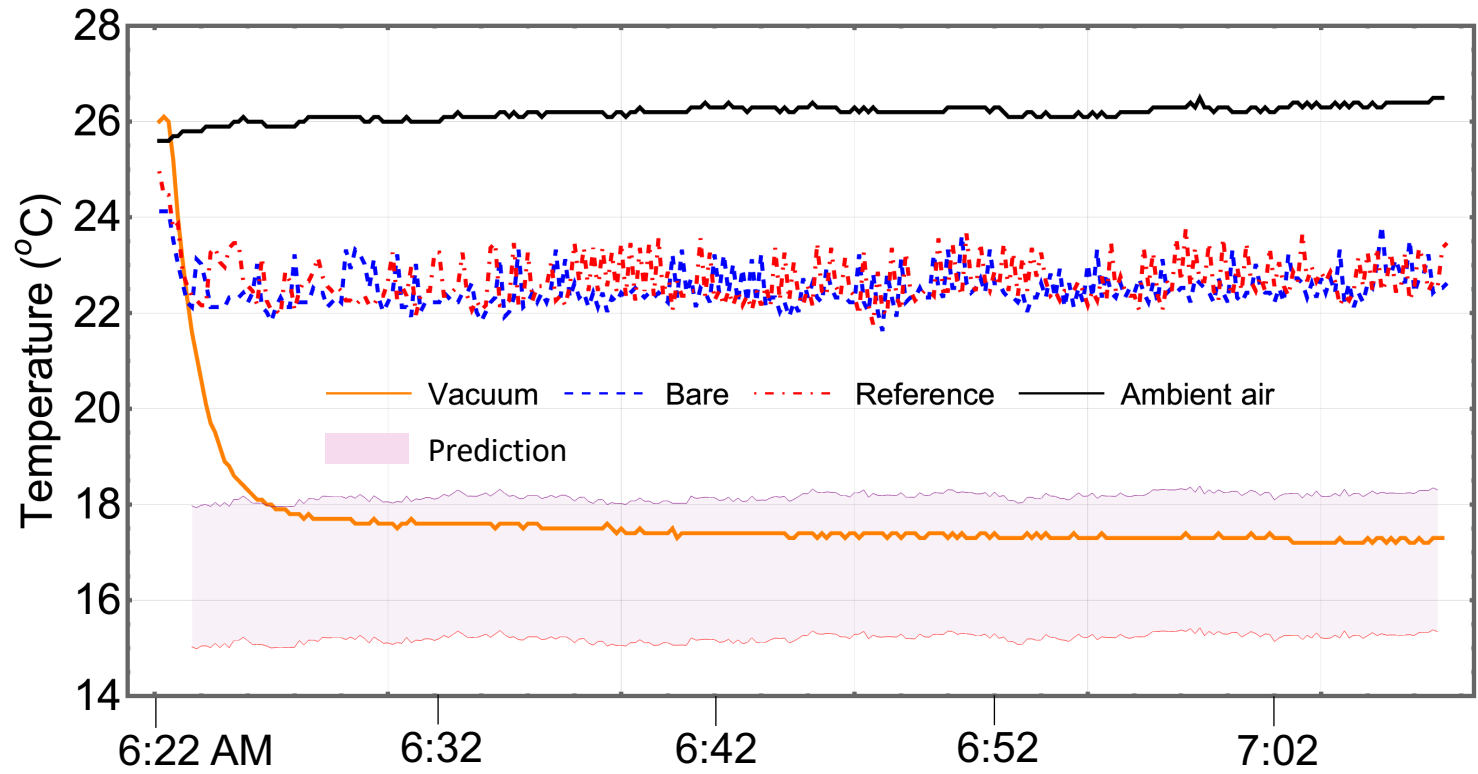

**Fig. 4b**

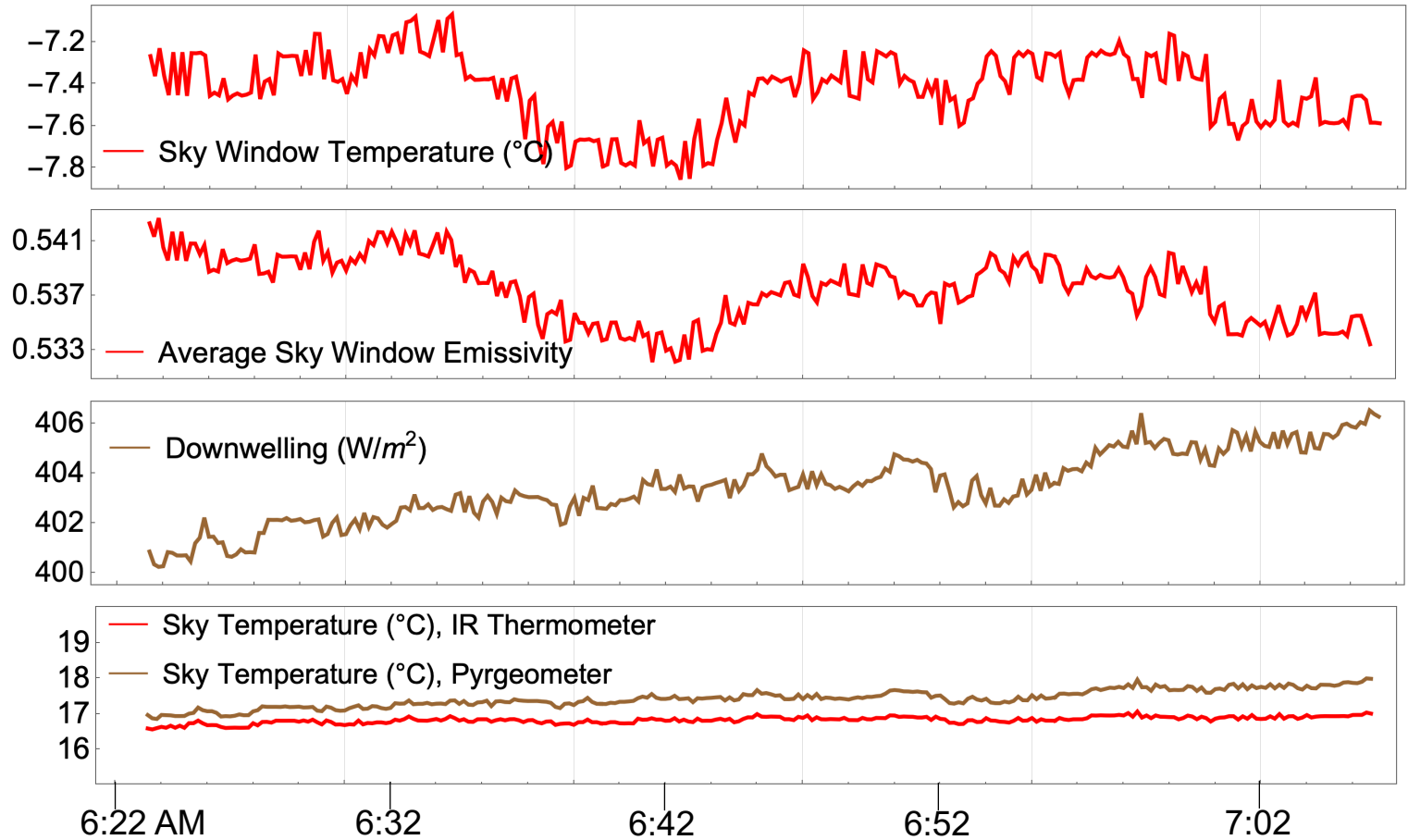

**Fig. 4c**

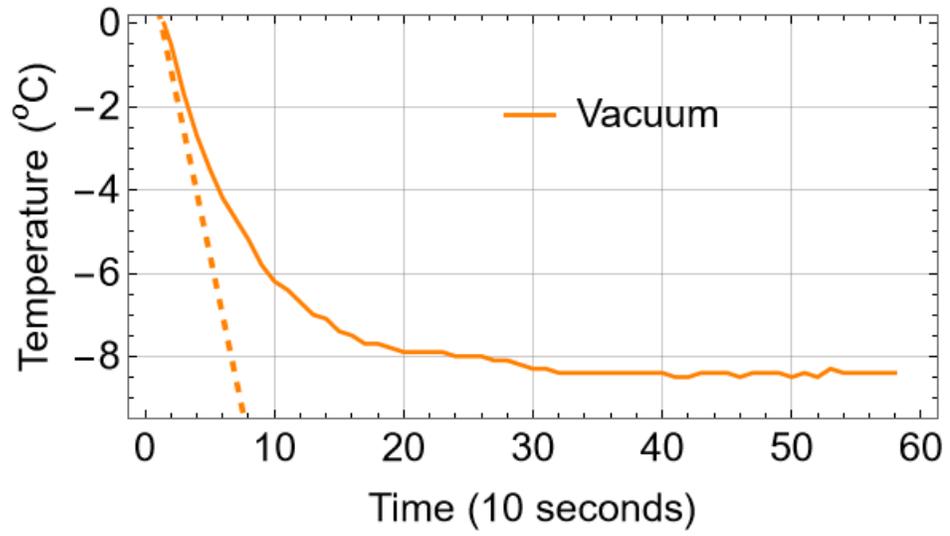

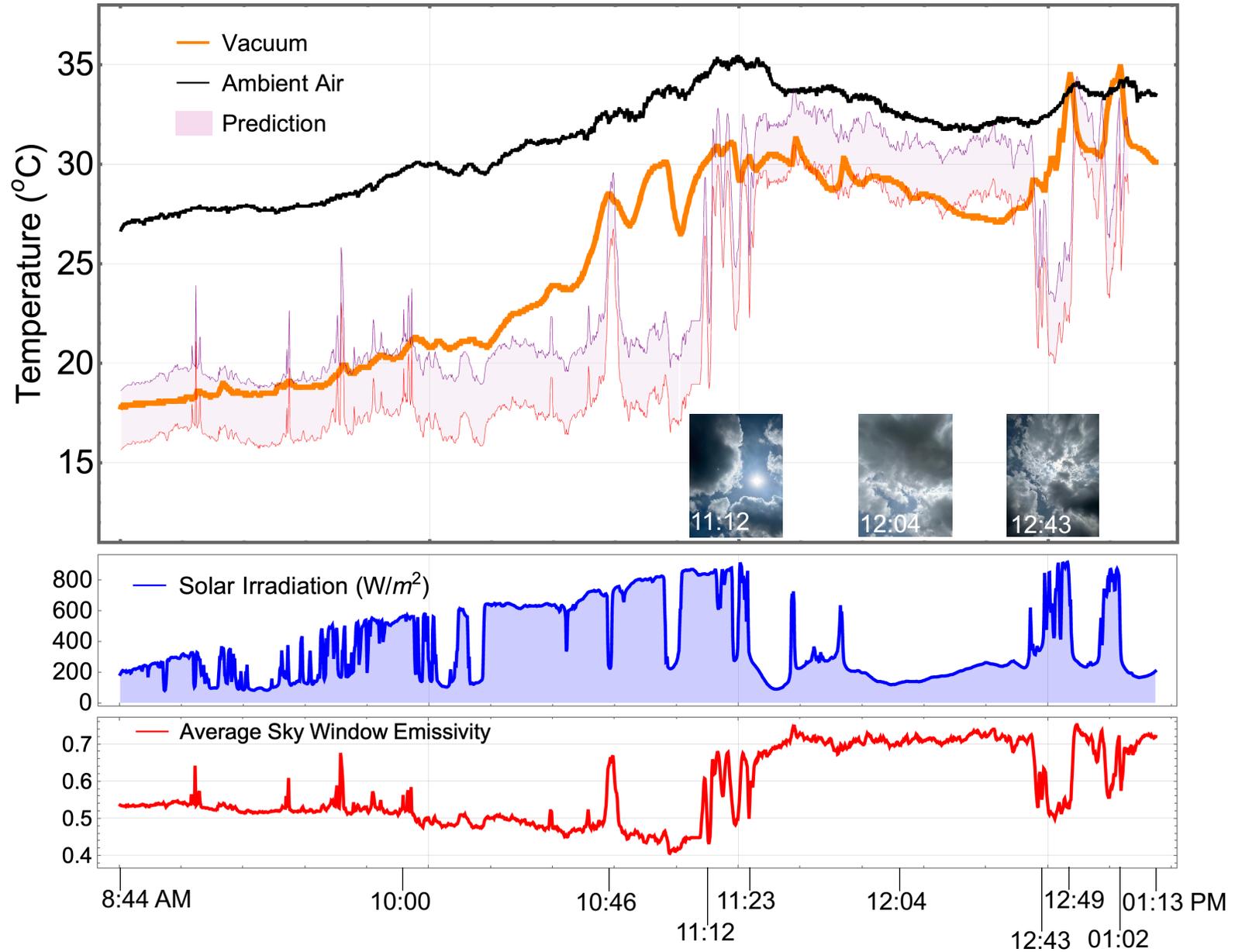

Fig. 5

**Supplementary Information**

## 1. Optical Characteristics

The transmission spectrum of the Germanium window was obtained with a Perkin Elmer Lambda 950 UV-Vis spectrometer in a transmission mode. The effective albedo $R$ of 0.998 is obtained from the blue overall transmission curve $T(\lambda)$ shown in Fig. 2b by $R = 1 - \int d\lambda S(\lambda) T(\lambda) / \int d\lambda S(\lambda)$, where $S(\lambda)$ is the solar spectrum.

Fig. S1 shows the infrared spectrum of the radiative cooling surface and the Germanium window. The total reflection spectrum of radiative cooling surface, 3M specular reflector attached to 3M aluminium foil tape, is presented. The total reflection spectrum was taken with an FT-IR spectrometer (Thermo Scientific, Nicolet iS-50), with the excitation at 12 degrees incidence angle, with an integrating sphere (Mid-IR IntegratIR, PIKE technology) to integrate both specular and diffuse components. The transmission spectrum at normal incidence is shown as provided by the ventor (Knight Optical). Germanium window is AR-coated with Diamond-Like Carbon (DLC). For guidance, the MODTRAN simulation result for the transmission of the atmosphere is shown [1]. The pre-set locality of "Tropical Atmosphere" was used with the following parameters: '$CO_2$ 400ppm, $CH_4$ 1.7ppm, Troposphere Ozone 28 ppb, Altitude 0 km looking up, Stratosphere Ozone scale 1, Water vapour scale 1, Freon scale 1, No clouds or Rain.'

In relation to the emissivity of the radiative cooling surface, the angle-dependent emissivity spectrum $e_s(\eta, \lambda)$ of the specular reflector published in ref [2] was used for the analyses, including the estimation of the cooling power and the lowest achievable temperature. The hemispherical emissivity $\varepsilon$ = 0.60 was estimated at $T_a$ = 300K from

$$\varepsilon = \int_0^{\pi/2} d(sin^2\eta) \int_0^\infty d\lambda P(\lambda, T_a) e_s(\eta, \lambda) / \sigma T_a^4. \quad (S1)$$

Similarly, the sky window emissivity $\varepsilon_{SW}$ = 0.79 was estimated at $T_a$ = 300K from

$$\varepsilon_{SW} = \int_0^{\pi/2} d(sin^2\eta) \int_{SW} d\lambda P(\lambda, T_a) e_s(\eta, \lambda) / \int_{SW} d\lambda P(\lambda, T_a), \quad (S2)$$

where SW represents the wavelength range of the sky window.

In relation to the reflection spectrum of the 3M aluminium tape in Fig. 2b, the total reflectance spectrum of the aluminium surface was obtained with a UV-VIS-IR optical spectrophotometer (Agilent Technologies, CARY-7000), with Diffuse Reflectance Accessories (DRAs) to integrate both specular and diffuse components.

## 2. Measurement Setup

The first panel of Fig. S2 shows the photo of the vacuum chamber. The lids, a top lid and a bottom lid, are connected to the vacuum chamber via KF flanges with Viton O-rings. A turbo pump and a vacuum gauge are also connected to with a KF flange with Viton O-rings. The lowest achievable pressure, $4.0 \times 10^{-5}$ mb, is deemed to be limited by the outgassing of the polymeric radiative cooling surface and the O-rings. The achievable base pressure could be improved by using CF flanges and using an inorganic radiative cooling surface, but since the heat transfer coefficient $h_{eff}$ is already negligible at that pressure, KF flanges were used for the ease of use. The Germanium window is seated inside an opening machined within the top lid to the vacuum chamber. The opening, with 9 cm diameter, has a lip with a groove to support a Viton O-ring (inner diameter: 9.2 cm, outer diameter 9.8 cm), directly in contact with the lower surface of the Germanium window.

A teflon structure was designed to support the radiative cooling surface and the radiation shields. The teflon structure included a pedestal matching the inner diameter of the vacuum chamber and was fitted at the bottom surface of the internal space of the chamber. The teflon pedestal had grooves to fix three vertical shields in cylindrical shapes and four pillars within the innermost vertical shields. The three vertical shields had 10.5 cm, 11 cm, 11. 5 cm diameter respectively and the horizontal shields had 10 cm diameter with 4 holes punched in to let the pillars through. The pillars were 8.5 cm apart between the opposing pillars, to position and support the horizontal shields. A teflon ring structure was fixed to the top tips of the teflon pillars, to which four strands of bare optical fibers were attached. The teflon surfaces were covered with aluminium foils to minimise the thermal radiation towards the radiative cooling surface.

The middle panel of Fig. S2 shows the measurement site with the radiative cooling setup. The vacuum chamber, the turbo pump and the measurement electronics were placed within an observatory building built for the operation of a space telescope. The observatory building was equipped with a roof that can be opened and closed in a controllable fashion. Since it rained at unpredictable times, the movable roof was used to protect the vacuum chamber, the equipment, and the measurement electronics in case of rain. The middle panel photo was taken when the roof was fully open. The vacuum chamber was placed at a height that ensures the view only to the sky through the concentrator structure without being obstructed by the walls of the observatory. The reference samples, the scotch tape sample and bare specular reflector sample were elevated to above the height of the wall of the observatory, to ensure a hemispherical exposure and to the ambient conditions of wind, temperature and humidity. During the day measurements, a heat spreader, which is 50cm by 50cm aluminium sheet covered with aluminised Mylar with an opening of 9 cm diameter was disposed on top surface of the Germanium window as shown in the photo. The Germanium window and the heat spreader was pushed down from the top with an aluminium ring with an indium seal in between.

The louver box to measure the ambient temperature was placed near the vacuum chamber, shielded from the direct solar illumination. A weather station (Davis Instruments 6250 Vantage Vue Wireless) was placed outside the observatory to monitor the wind speed, humidity and ambient temperature every minute. The ambient temperature measured by the weather station agreed well with that measured with the PT100 sensor in the louver box.

The offsets between the thermocouples and the PT100, less than ±0.3 degrees, were calibrated by measuring the temperatures for an hour after placing the sensors together in a box with multiple layers of thermal insulation. Data loggers (Calex Excelog 6 and Omega OM-CP-OCTPRO) were used to record the data from the temperature sensors, the pyrgeometer and the pyranometer every 10 seconds.

The rightmost panel of Fig. S3 shows view into the concentrator structure in use, fitted to the vacuum chamber around the Germanium window. The concentrator structure had a lower opening with a 10.5cm diameter and a higher opening with 34.5cm diameter. The height between the lower opening and the higher opening is 22cm. The outer surface of the concentrator was covered with aluminised Mylar.

The radiation shields, both horizontal and vertical shields and the concentrator structure were made with 400 microns thick polished aluminium sheets. The aluminium sheets were made by hand-polishing with cotton pads, first with WD40 and subsequently with Autosol polishing cream. Fig. S3 shows total and diffuse reflectance spectra taken with an FT-IR spectrometer (Thermo Scientific, Nicolet iS-50), with an integrating sphere (Mid-IR IntegratIR, PIKE technology) for integrating both specular and diffuse components. The reflectance of the range overlapping with the sky window 7 to 14 $\mu m$, is higher than 95%. The diffuse component of the reflection in the sky window is less than 10%.

## 3. Net cooling power

The analysis below is based on the formalism of ref [3], modified by including $P_3$. The net radiative cooling power $P_{rad}(T_a, T_s)$ as a function of the ambient temperature $T_a$ and the radiative cooling substrate temperature $T_s$ is given by:

$$P_{rad}(T_a, T_s) = P_1(T_s) - P_2(T_a) - P_3(T_a) - P_{parasitic}(T_a, T_s), \qquad (2)$$

$P_1$ is the power radiated from the radiative cooling substrate at temperature $T_s$:

$$P_1(T_s) = \int_0^{\pi/2} d(sin^2\eta) \int_0^\infty d\lambda P(\lambda, T_s) e_s(\eta, \lambda), \qquad (S3)$$

where $e_s(\eta, \lambda)$ is the angle-dependent emissivity spectrum of the radiative cooling surface and $P(\lambda, T_s)$ is the Planck spectrum. $P_2$ is the downwelling absorbed by the radiative cooling substrate within the acceptance angle $\eta_{max}$ of the concentrator structure:

$$P_2(T_a) = \int_0^{\eta_{max}} d(sin^2\eta) \int_0^\infty d\lambda P(\lambda, T_a) e_a(\eta, \lambda) e_s(\eta, \lambda) =$$
$$\int_0^{\pi/2} d(sin^2\eta) \left( \int_{SW} d\lambda P(\lambda, T_a)(1 - (1 - e_{avg,sw})^{1/\cos\eta}) e_s(\eta, \lambda) + \int_{NSW} d\lambda P(\lambda, T_a) e_s(\eta, \lambda) \right), \qquad (S4)$$

where SW represents the wavelength range of the sky window and NSW represents the wavelength range outside the sky window. $P_3$ is the "reciprocal rays" term, which accounts for the downwelling in the zenith direction incident on the radiative cooling surface by being reflected by the concentrator structure and absorbed by the radiative cooling surface:

$$P_3(T_a) = \int_{\eta_{max}}^{\pi/2} d(sin^2\eta) \int_0^\infty d\lambda P(\lambda, T_a) e_a(0, \lambda) e_s(\eta, \lambda) =$$
$$\int_{\eta_{max}}^{\pi/2} d(sin^2\eta) \left( \int_{SW} d\lambda P(\lambda, T_a) e_{avg,sw} e_s(\eta, \lambda) + \int_{NSW} d\lambda P(\lambda, T_a) e_s(\eta, \lambda) \right). \qquad (S5)$$

It is noted that it is assumed for $P_2(T_a)$ and $P_3(T_a)$ that the concentrator structure is designed to direct all of the rays from the radiative cooling substrate to the zenith direction with emissivity $e_a(0, \lambda)$. For example, the concentrator structure can be a parabolic mirror with the radiative cooling substrate at the focus of the parabola. Since the concentrating structure is of a truncated cone shape, the rays from the radiative cooling substrate are generally directed in the zenith direction but with a finite angular distribution. However, the above equation for net cooling power $P_{rad}$ is a good approximation for our system because due to the opaque sky of the tropical climate, the emissivity varies slowly with angle from the zenith direction. For example, with an opaque sky window with average sky window emissivity $e_{avg,sw} = 0.6$, the emissivity in the angle $\frac{\pi}{10}$ from the zenith direction, $e_a\left(\frac{\pi}{10}, \lambda\right)$, is 5 % smaller than the emissivity in the zenith direction $e_a(0, \lambda)$.

The term "cooling power" is used to represent the initial cooling power $P_{rad}$ at ambient temperature when $T_a = T_s$, as conventionally used in the radiative cooling literatures. As discussed in Fig. 4c of the main text, we estimated the cooling power by starting radiative cooling abruptly and measuring the slope of the temperature trajectory. Fig. S4 shows the cooling power values obtained under a clear sky at each sky window temperature $T_{SW}$ at the time of the measurement. A general trend of a larger cooling power for a colder sky window is observed.

## 4. Calibration measurements

In relation to the calibration measurements with ice presented in Fig. 3, the net cooling power is written as:

$$P_{rad}(T_s) = P_1(T_s) - P_2(T_{ice}) - h_{eff}(T_a - T_s), \text{ where}$$
$$P_2(T_{ice}) = \int_0^{\pi/2} d(sin^2\eta) \int_0^\infty d\lambda P(\lambda, T_{ice}) e_s(\eta, \lambda). \quad (S6)$$

The first term $P_1$ is as given in equation (S3). The second term $P_2(T_{ice})$ accounts for the infrared radiation from the ice pack incident on the radiative cooling substrate. Since the ice pack is assumed to be a blackbody at 0 degrees, the second term $P_2(T_{ice})$ corresponds to $P_2$ of equation (S4) with $\eta_{max} = \pi/2$, corresponding to the hemispherical acceptance, $T_{ice} = 273K$ and $e_{ice}(\eta, \lambda) = 1$ over all wavelengths. There is no power contribution corresponding to $P_3$ of equation (S5) because no concentrator structure is used. As discussed in the main text, at $4.5 \times 10^{-5}$ mb pressure, the parasitic heat gain can be ignored such that $h_{eff} \sim 0$. The temperature reduction at the steady state temperature is obtained by solving (S6) $P_1(T_s) - P_2(T_{ice}) = 0$. The lowest steady state temperature calculated according to equation (S6) is 26 degrees, much lower than observed 14 degrees. This is attributed to multiple factors: the ice pack does not covering the entire hemisphere over the radiative cooling surface, the finite transmission of the Germanium window and most importantly radiative warming due to the finite emissivity of the internal surface of the Germanium window and the radiation shields. Therefore, a modification of equations (2) and (S6), taking these factors in consideration, is necessary. Taking a cue from the fact that these factors are largely independent of the radiative cooling surface temperature, the discrepancy with measurement data is calibrated with a single term, $P_{cal}$. The cooling power of the calibration measurements is modified to:

$$P_{rad}(T_s) = P_1(T_s) - P_2(T_{ice}) - P_{cal}. \quad (S6')$$

The same calibration power $P_{cal}$ is also included in equation (2'). To estimate the calibration term $P_{cal}$, an additional surface at ambient temperature, $T_a$, is assumed to be present in radiative exchange with the radiative cooling surface. The corresponding power of the calibration contribution is formulated as:

$$P_{cal}(T_a, \alpha) = \alpha \int_0^{\pi/2} d(sin^2\eta) \int_0^\infty d\lambda P(\lambda, T_a) e_s(\eta, \lambda), \quad (S7)$$

where $\alpha$ represents the average effective emissivity of the surrounding surfaces. The calibration power $P_{cal}$ can be estimated from the time trajectory of radiative cooling surface temperature. We used two points in the trajectory: $P_{rad}(T_a, T_{s\_min}) = 0$, where $T_{s\_min}$ is the steady state minimum temperature and $P_{rad}(T_a, T_s = T_a) = P_{rad}$, where $P_{rad}$ is the initial cooling power as evaluated in Fig. 4c. The effective emissivity $\alpha$ values were measured and collected from the calibration measurement using equation (S6') and also in the field measurements using equation (2'). The effective emissivity $\alpha$ ranged from 0.11 to 0.15. The shaded area of Fig. 4a and Fig. 5 represents the uncertainty corresponding to the range of the effective emissivity $\alpha$.

To confirm the characteristic length $d$ of the relation between the pressure and thermal conductivity of air, $\frac{k}{k_0} = 1/\left(1 + \frac{CT}{pd}\right)$ is the distance between the radiative cooling surface and the inside-facing surface of the Germanium window, measurements similar to Fig. 3 were performed after the distance $d$ was increased to 11 cm, as shown in Fig. S5. The radiative cooling surface and the horizontal shields were lowered by 10.2 cm. The vertical shields in cylindrical shapes were not moved. Changing the distance $d$ lowers the magnitude of the temperature reduction and changes the pressure dependence from that presented in Fig. 3, although the general trend of improvement with lower pressure persists. Notably, the scroll pump pressure 8 mbar slightly improves the temperature reduction with respect to the atmospheric pressure in the chamber. The largest improvement is seen from $1.9 \times 10^{-2}$ mbar pressure to $2.0 \times 10^{-3}$ mbar pressure. The cooling performance did not improve significantly from $2.0 \times 10^{-3}$ mbar pressure to $2.0 \times 10^{-4}$ mbar pressure, although there is a noticeable improvement from $2.0 \times 10^{-4}$ mbar pressure to o $4.5 \times 10^{-5}$ mbar pressure. These complex behaviours are attributed to the radiative exchange with other surfaces at a closer distance than $d$, such as the surfaces of the innermost vertical shield. This measurement confirms that the pressure dependence changes drastically when distance $d$ between the lower surface of the Germanium window and the radiative cooling surface.

To measure the effectiveness of radiation shields, temperatures at various positions within the chamber were monitored during the icepack measurement. Fig. S6 shows the calibration measurement made with icepack in Fig. 3, only the curve at $4.5 \times 10^{-5}$ mbar pressure with the time duration stretching beyond that presented in Fig. 3. At around 1500 seconds, the window is blocked with aluminium plate to stop radiative cooling, from which the black curve, the temperature of the radiative cooling surface, increases. Simultaneous measurements with additional K-type thermocouples placed within the vacuum chamber are shown: in the space outside the outermost vertical shield (orange curve), in the space below the lowest horizontal shield (blue curve), disposed on the bottom of the lowest horizontal shield (purple curve). Both the temperature sensor on and below the lowest horizontal shield do not react to the presence or the absence of the ice pack. This shows that the number of the horizontal shields is sufficient to suppress the radiative exchange in vertical direction. In contrast, the space outside the third and outermost layer of the vertical shields cools down by around 0.5 degree during 1500 seconds of cooling. When the window is blocked, this space tends to warm back up, but much slower than the radiative cooling surface. This suggests that there is non-negligible amount of sideways radiative exchange, not completely suppressed by three layers of radiative shields. However, since the radiative cooling surface lies horizontally has a negligible side-facing surface area, this sideways radiative exchange is considered not to affect the performance significantly.

## 5. Sky window temperature and sky window emissivity

*Infrared thermometer and sky window temperature*

We directly measure the sky window emissivity $e_{avg,sw}$ by pointing an infrared thermometer in the zenith direction and use the measured sky window emissivity value $e_{avg,sw}$ for the atmospheric emissivity model in equation (1). We used an IR pyrocouple (Calex PC151LT-0mA) with an active response in the 8 to 14 $\mu m$ wavelength range, a fixed emissivity of $\varepsilon_{IR} = 0.95$ and a field of view 15:1. The infrared emission power received by the IR thermometer corresponds to the $P_2$ of equation (S4) with $\eta_{max} = 1/15$ and $e_{s,sw}(\eta, \lambda) = \varepsilon_{IR} = 0.95$ and $e_{s,nsw}(\eta, \lambda) = 0$.

$$P_{IRTherm} = \varepsilon_{IR} \int_0^{\eta_{max}} d(sin^2\eta) \int_0^\infty d\lambda P(\lambda, T_a) e_a(\eta, \lambda) e_s(\eta, \lambda)$$
$$= \varepsilon_{IR} \int_0^{\eta_{max}} d(sin^2\eta) \int_{SW} d\lambda P(\lambda, T_a) \left(1 - (1 - e_{avg,sw})^{\frac{1}{\cos\eta}}\right)$$
$$\propto \varepsilon_{IR} \int_{SW} d\lambda P(\lambda, T_a) \, e_{avg,sw}. \text{ (S8)}$$

The last approximation is based on that when $\eta_{max} = 1/15$, $\int_0^{\eta_{max}} d(sin^2\eta) \sim \eta_{max}^2$ and $1/\cos\eta_{max} \sim 1$. The power detected by the infrared thermometer is therefore directly proportional to the sky window emissivity in the zenith direction, $e_{avg,sw}$. In practice, the IR thermometer directly outputs a current signal corresponding to the sky window temperature $T_{SW}$, which is calibrated to the received infrared power within the 8 to 14 $\mu m$ wavelength range of $\varepsilon_{IR} P(\lambda, T_{SW})$. The sky window emissivity in the zenith direction, $e_{avg,sw}$ can be obtained from the sky window temperature $T_{SW}$ with $e_{avg,sw} = \int_{SW} d\lambda \varepsilon_{IR} P(\lambda, T_{SW})/\int_{SW} P(\lambda, T_a) d\lambda$.

*Pyrgeometer and sky temperature*

The sky temperature $T_{sky}$ is defined as $P_{Pyrgeo} = \sigma T_{sky}^4$, where $P_{Pyrgeo}$ is the atmospheric emission over the hemisphere and can be measured with a Pyrgeometer. The power measured by a Pyrgeometer corresponds to $P_2$ of equation (S4) with the hemispherical acceptance $\eta_{max} = \pi/2$, assuming the receiving substrate to be a black body, $e_s(\eta, \lambda) = 1$.

$$P_{Pyrgeo} = \int_0^{\pi/2} d(sin^2\eta) \int_0^\infty d\lambda P(\lambda, T_a) e_a(\eta, \lambda) e_s(\eta, \lambda)$$

$$= \int_0^{\pi/2} d(sin^2\eta) \int_0^\infty d\lambda P(\lambda, T_a) e_a(\eta, \lambda)$$

$$= \int_0^{\pi/2} d(sin^2\eta) \left( \int_{SW} d\lambda P(\lambda, T_a) \left(1 - (1 - e_{avg,sw})^{\frac{1}{\cos\eta}}\right) + \int_{NSW} d\lambda P(\lambda, T_a) \right)$$

$$= \sigma T_a^4 - \int_0^{\pi/2} d(sin^2\eta) \left( \int_{SW} d\lambda P(\lambda, T_a) (1 - e_{avg,sw})^{\frac{1}{\cos\eta}} \right) \quad . \quad (S9)$$

The thermopile inside a Pyrgeometer measures a quantity corresponding to the second term of the last line of equation (S9). The second term also represents the long wave imbalance, the net difference between the blackbody emission at the ambient temperature $T_a$ and the atmospheric emission. It is noted that the definition of the sky temperature, $P_{Pyrgeo} = \sigma T_{sky}^4$, is based on the assumption that the overall atmospheric emission follows a Planck spectrum with the total power $P_{Pyrgeo}$. The atmospheric model of equation (1) is used to better reflect the actual spectrum without such assumption. The sky window temperature is directly related to the sky temperature only if the sky condition is uniform throughout the hemisphere. In this case, the average sky window emissivity $e_{avg,sw}$ can be estimated with the measurement of the sky temperature, $T_{sky}$. For a cloudy sky, the sky window temperature $T_{SW}$ is more indicative for radiative cooling than the conventional sky temperature $T_{sky}$.

*Atmospheric model, sky temperature and sky window temperature*

The sky window temperature, $T_{SW}$, coined in this work, is to be differentiated with the conventionally defined "sky temperature," $T_{sky}$. Fig. S7 is an example that illustrates the model atmosphere according to equation (1) and the relationship between the sky window temperature $T_{SW}$ and the conventionally defined sky temperature $T_{sky}$. The example relates to the MODTRAN simulation results [1] with the pre-set locality of "Tropical Atmosphere" with the same parameters disclosed in Fig. S1. The purple solid line is the MODTRAN simulation result for the atmospheric radiation towards the earth at 26.7 °C ambient temperature or $T_a$ = 299.7 K. The Planck spectrum at the ambient temperature is shown as the black dotted curve. Outside the sky window of 8 to 14 $\mu m$ wavelength range, the MODTRAN atmospheric spectrum closely follows the Planck spectrum. The ratio of the atmospheric emission to the Planck spectrum is shown as the purple solid line in the right panel. As discussed in the equation (S8), the infrared thermometer measures the emission power over the sky window and outputs a corresponding temperature. For the atmosphere represented by the MODTRAN simulation results presented in Fig. S7, the infrared thermometer pointing towards the zenith direction would measure -8.5 °C, in other words, the sky window temperature, $T_{SW}$, of 264.5 K. The Planck spectrum at this temperature is shown as the blue dotted line in the left panel. The atmospheric model according to equation (1) constructed therefrom is shown as the red solid line, both in the left and right panels. In the left panel, the atmospheric emission spectrum follows the Planck spectrum at the ambient temperature $P(\lambda, T_a)$ outside the sky window, and the Planck spectrum at the sky window temperature $P(\lambda, T_{SW})$ within the sky window. In the right panel, the emissivity spectrum of the atmosphere is $e_{avg,sw}$, a single measured value within the sky window and unity outside the sky window. The average sky window emissivity $e_{avg,sw}$ according to the MODTRAN simulation is 0.52 as shown in the right panel of Fig. S8.

Assuming a uniform sky, the total hemispherical atmospheric emission $P_{Pyrgeo}$ and the sky temperature $T_{sky}$ evaluated based on this atmospheric model according to equation (S9) is around 16.9 °C, $T_{sky}$= 289.9 K. The Planck spectrum at the sky temperature $P(\lambda, T_{sky})$ is shown in the left panel as a brown solid line. As discussed above, the sky temperature $T_{sky}$ is defined based on the assumption that the atmospheric spectrum follows a Planck distribution. Therefore, the area under the atmospheric spectrum, either from the MODTRAN simulations (purple solid line) or from the atmospheric model of equation (1) (red solid line), equals the area under the Planck spectrum at the sky temperature $P(\lambda, T_{sky})$. As a result, the ambient temperature $T_a$ is higher than both the sky temperature $T_{sky}$ and the sky window temperature $T_{SW}$. The sky window temperature $T_{SW}$ is lower than the sky temperature $T_{sky}$.

## 6. Daytime radiative cooling measurements

*The influence of solar heating*

Fig. S8 shows the concurrent measurements performed with those presented in Fig. 4 of the main text. The orange line, the radiative cooling surface temperature in the vacuum chamber, and the black line, the ambient air, are as in Fig. 4 and reproduced here for reference. Additionally shown in Fig. S8 are the temperature of the bottom shield (blue line), the temperature on the side surface of the Germanium window (red line) and the temperature in the space outside the outermost vertical shield (purple line). The temperature outside the outermost vertical shield generally follows the ambient temperature. As expected, the Germanium window heats above the ambient temperature when the solar irradiation level is high, as shown in the pyranometer measurements, reproduced from Fig. 5, notwithstanding the use of the heat spreader. The radiative cooling surface temperature is generally correlated with the temperature of the Germanium window. Whenever the temperature of the Germanium window rises above the ambient temperature, the temperature reduction decreases dramatically. Since the temperature sensor was attached to the side surface of the Germanium window, the temperature towards the middle of the Germanium window is expected to be higher than the measured temperature presented in Fig. S8. The bottom shield temperature is largely immune to the trend of the solar irradiation, only showing the slow increasing trend, staying below the ambient temperature. This behaviour is attributed to the heat capacity of the radiation shield itself.

*Reference samples temperature*

Fig. S9 shows the concurrent measurements performed with those presented in Fig. 4 of the main text. The orange line, the radiative cooling surface temperature in the vacuum chamber, and the black line, the ambient air, are as in Fig. 4 and reproduced here for reference. Additionally shown in Fig. S10 are the temperature of the scotch tape sample (red line) and the temperature of the bare specular reflector sample or the reference sample (purple line). See the main text for Fig. 3a regarding the configurations of these samples. The scotch tape sample heats significantly above the ambient temperature when the solar irradiation level is high (see Pyranometer measurements of Fig. 4 in the main text). The bare specular reflector sample, identical to the radiative cooling surface in the vacuum chamber occasional cools below and heats above the ambient temperature by a couple of degrees depending on the level of solar irradiation. Considering that two samples showed similar cooling performance in absence of sunlight, the superior performance of the specular reflector specular reflector on aluminium tape is attributed to the higher solar reflectivity. This shows that a simple surface can mitigate severe overheating under direct sunlight but is not capable of staying below the ambient temperature in Singapore.

## 7. Estimation of cooling power and steady state temperature.

Fig. S10 is the cooling power plotted as a function of the temperature reduction from the ambient temperature according to equation (2') in the main text and equation (S7). The specular reflector used in this work is assumed to be the radiative cooling surface and the ambient temperature is taken to be 300 K. Two representative values of the sky window emissivity $e_{avg,sw}$ are used, 0.53 and 0.7, for a clear sky and a cloudy sky in Singapore, respectively. The parasitic heat gain is considered suppressed with a high vacuum chamber such that $h_{eff} \sim 0$. The finite emissivity of the vacuum chamber is taken into consideration with $\alpha$ = 0.11. An ideal vacuum chamber with a perfectly transmissive window and non-emissive internal surfaces, where $\alpha$ = 0, is considered for the theoretical prediction of the best-case performance. Under a clear sky of Singapore, 62 $W/m^2$ cooling power and 19 degrees temperature reduction are achievable. Under a cloudy sky, 38 $W/m^2$ cooling power and 11 degrees temperature reduction are predicted. Notably, for the cloudy sky, the cooling power is improved by more than a factor of two when an ideal vacuum chamber is used.

## References

[1] D. Archer, MODTRAN, http://climatemodels.uchicago.edu/modtran/, University of Chicago.
[2] A.R. Gentle, G.B. Smith, A Subambient Open Roof Surface under the Mid-Summer Sun, Advanced Science, 2 (2015).
[3] G.B. Smith, Amplified radiative cooling via optimised combinations of aperture geometry and spectral emittance profiles of surfaces and the atmosphere, Solar Energy Materials and Solar Cells, 93 (2009) 1696-1701.

Fig. S1. Infrared spectra of the radiative cooling surface used in the experiment (orange), the Germanium window (purple). The transmission of the atmosphere in the tropical weather obtained by MODTRAN simulation is shown (gray) for reference.

Fig. S2 Left: the vacuum-enhanced radiative cooling setup. Middle: the measurement site, with the radiative cooling setup in the observatory building with the roof fully open. Right: the inner surface of the concentrator structure and the top surface of the Germanium window.

Fig. S3. The total and diffuse reflection spectra of polished aluminium surface.

Fig. S4. Cooling power at different sky window temperatures.

Fig. S5. The repetition of the ice pack measurement of Fig. 3 with $d = 11$ cm.

Fig. S6. Temperatures at the various positions within the chamber during the ice pack measurement.

Fig. S7. left: the MODTRAN simulation of atmospheric spectrum of the tropical atmosphere (purple) and the model atmosphere discussed in this work (red). The Planck spectrum at the ambient temperature (black), the sky temperature (brown) and the sky window temperature (blue) are shown. Right: the emissivity spectrum of the atmosphere model (red) shown with the simulation result normalised with the Planck spectrum at the ambient temperature (purple).

Fig. S8. Simultaneous measurements performed with those presented in Fig. 4: the temperature of the bottom shield (blue line), the temperature on the side surface of the Germanium window (red line) and the temperature in the space outside the outermost vertical shield (purple line).

Fig. S9. Simultaneous measurements performed with those presented in Fig. 4: the temperature of the scotch tape sample (red line) and the temperature of the bare specular reflector sample or the reference sample (purple line).

Fig. S10. Cooling power as a function of the temperature reduction for representative emissivity values for a clear sky and cloudy sky in Singapore and for an ideal vacuum chamber and a realistic vacuum chamber.

**Fig. S1**

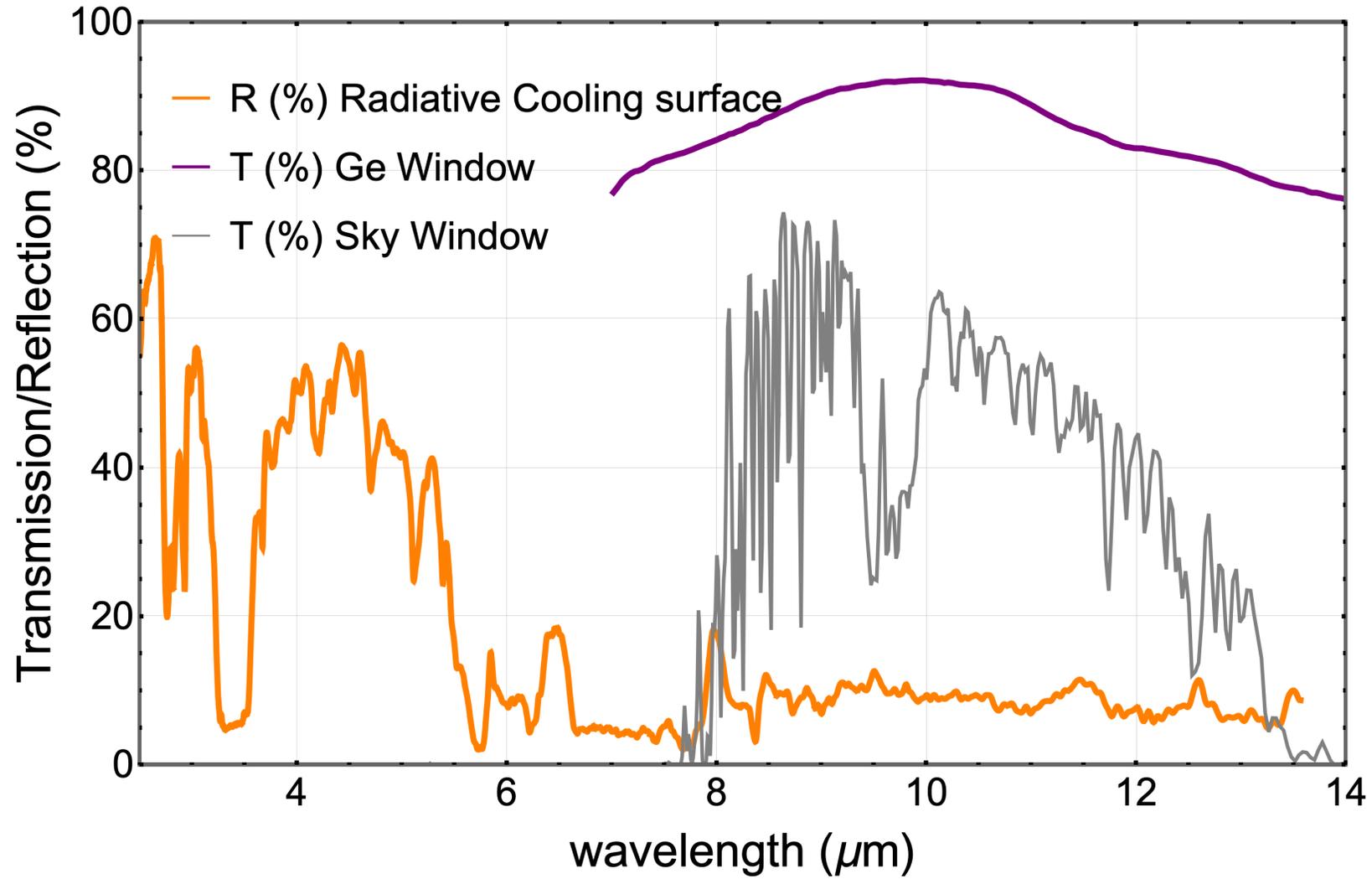

**Fig. S2**

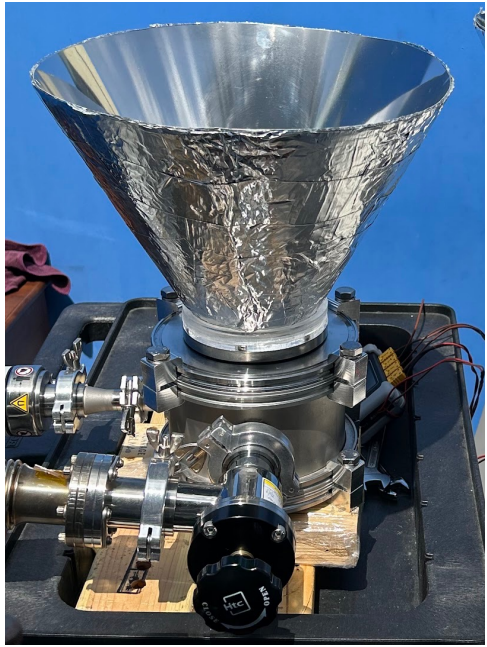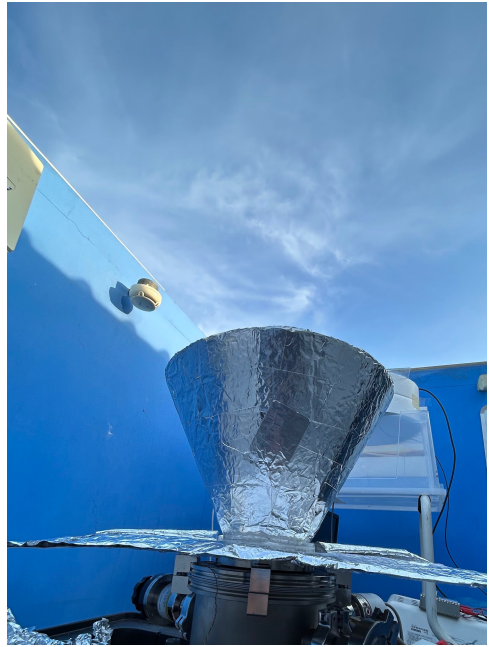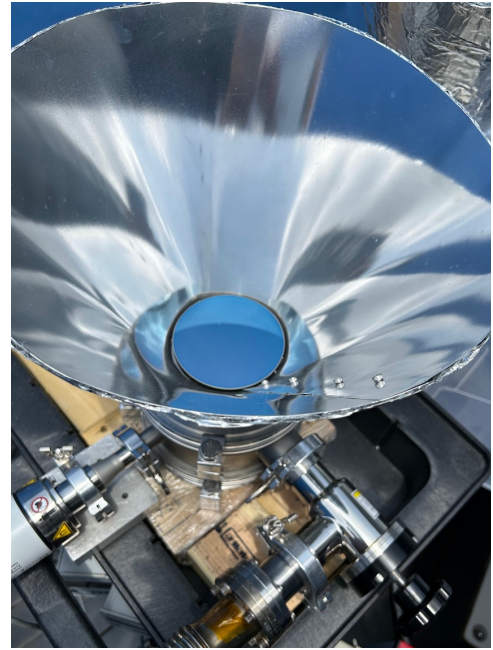

**Fig. S3**

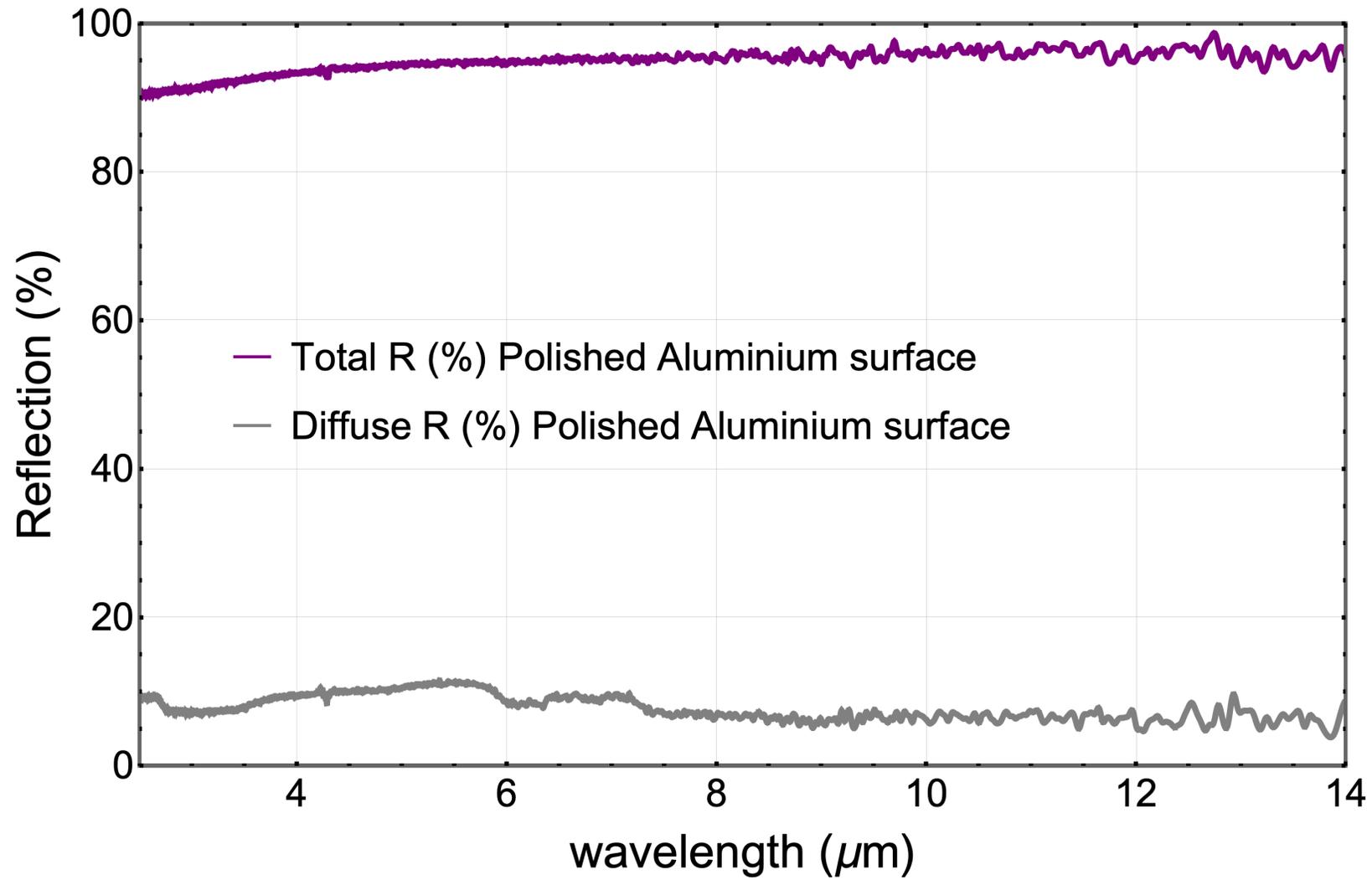

**Fig. S4**

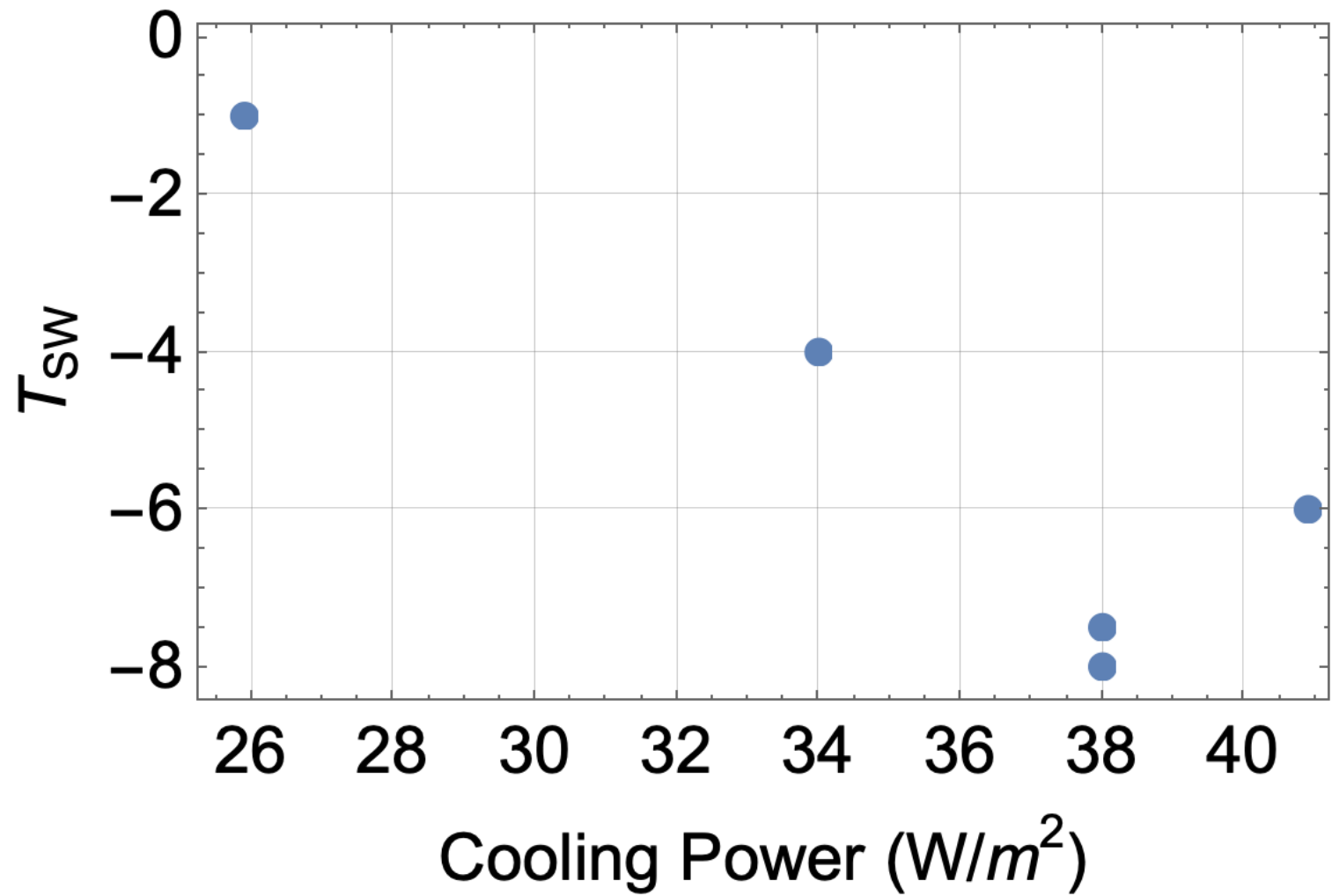

**Fig. S5**

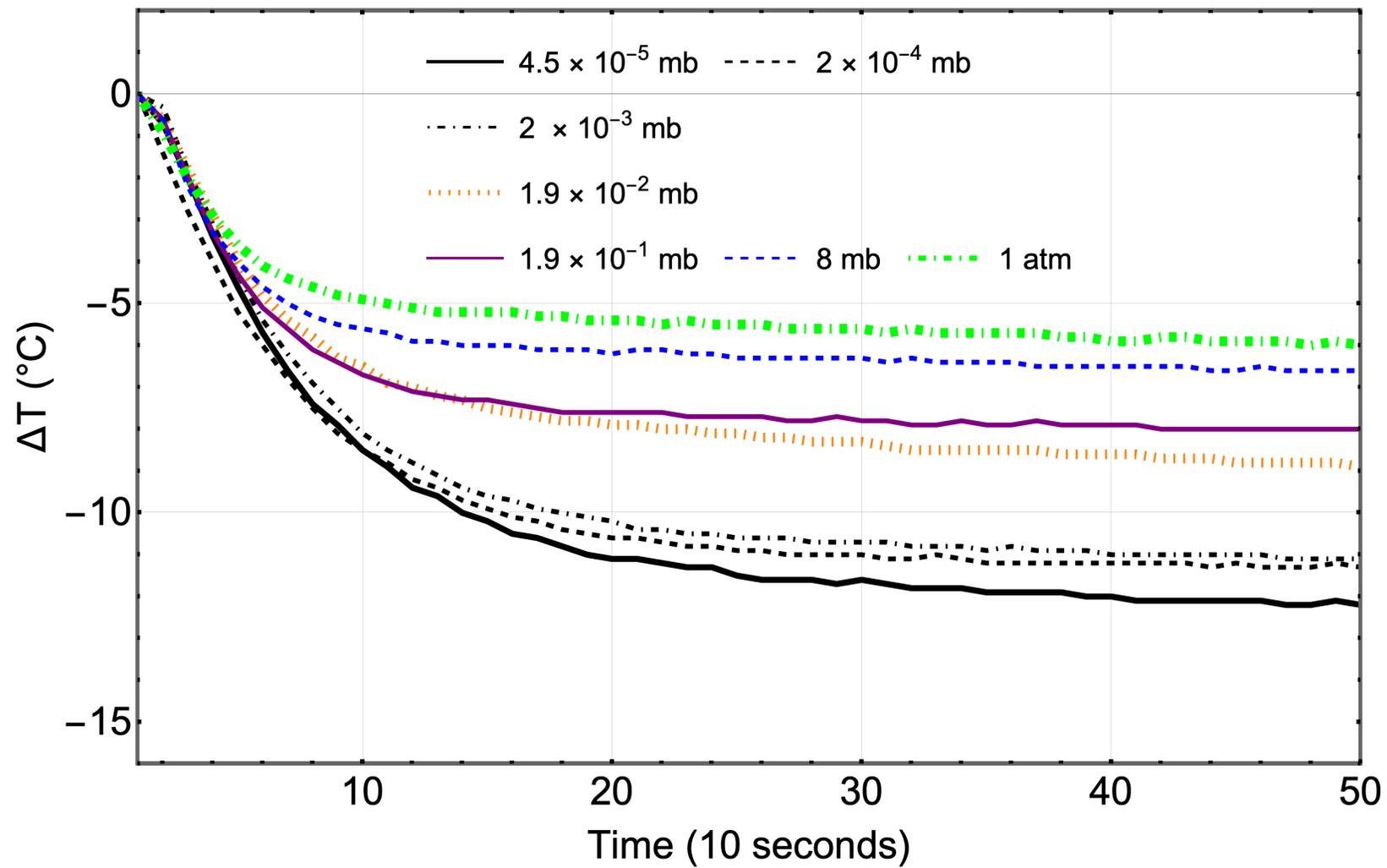

**Fig. S6**

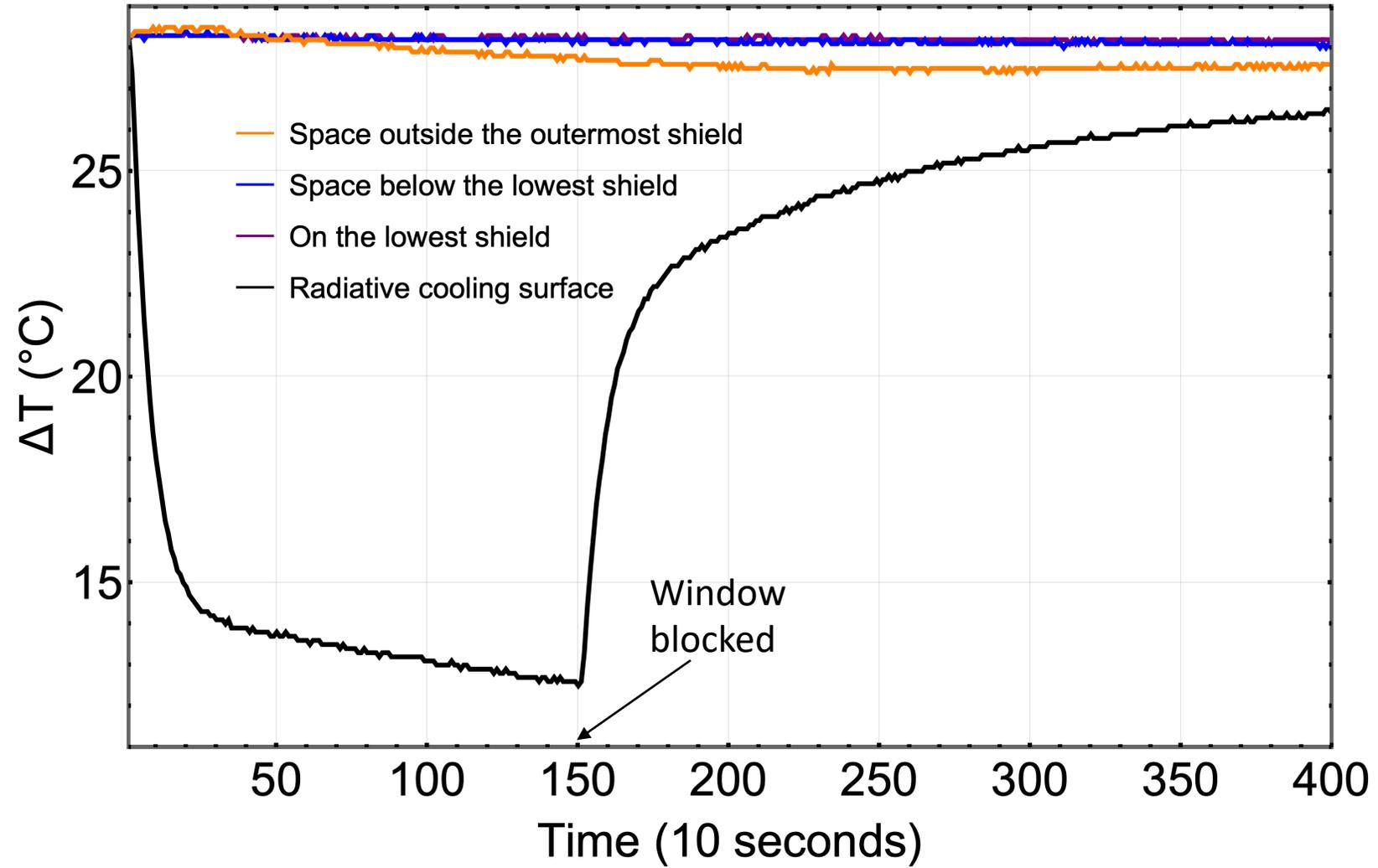

**Fig. S7**

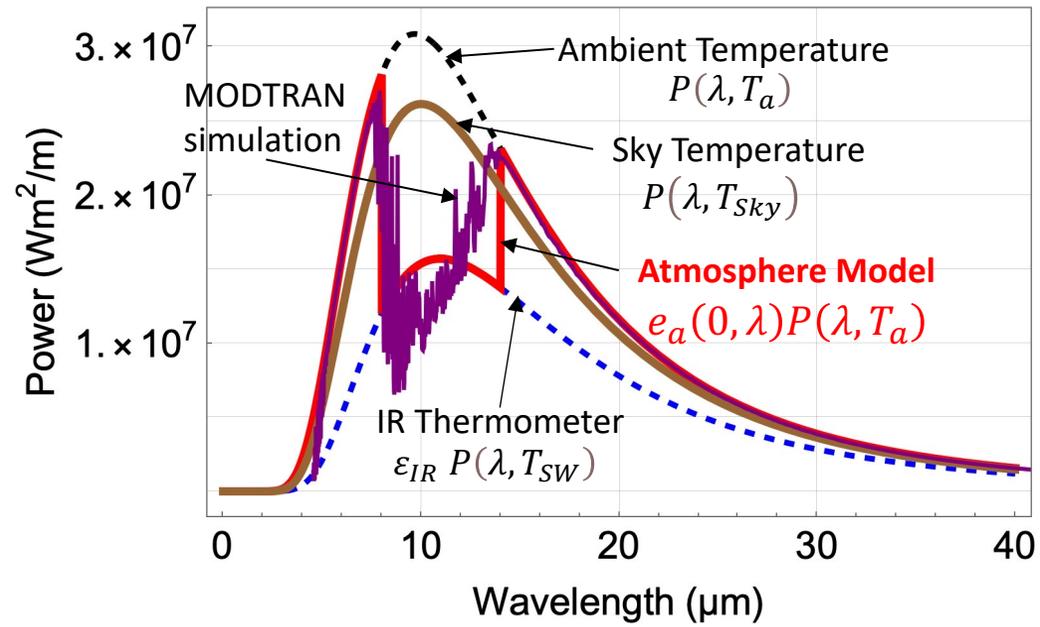 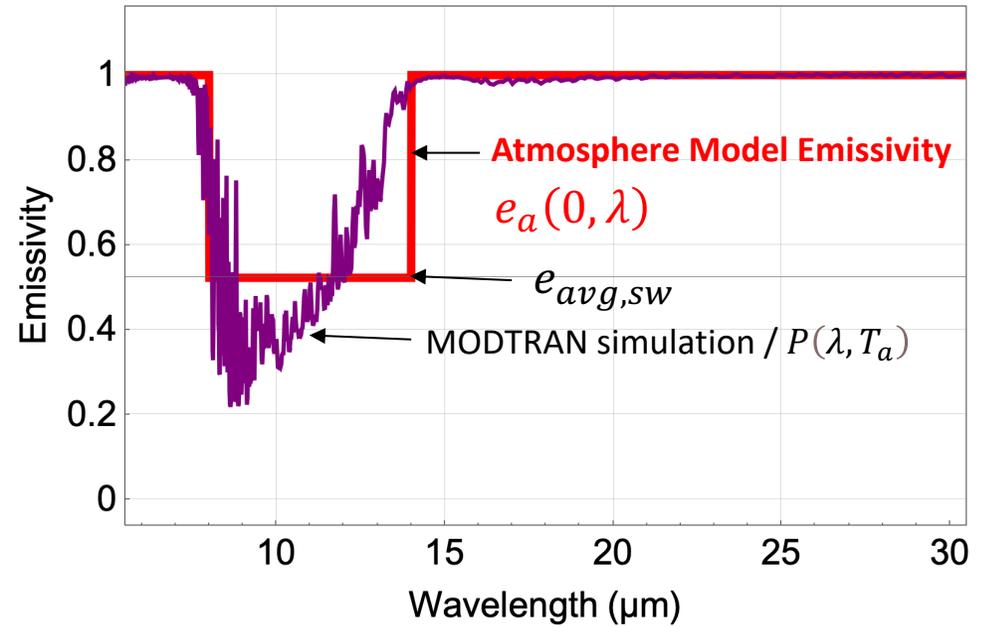

**Fig. S8**

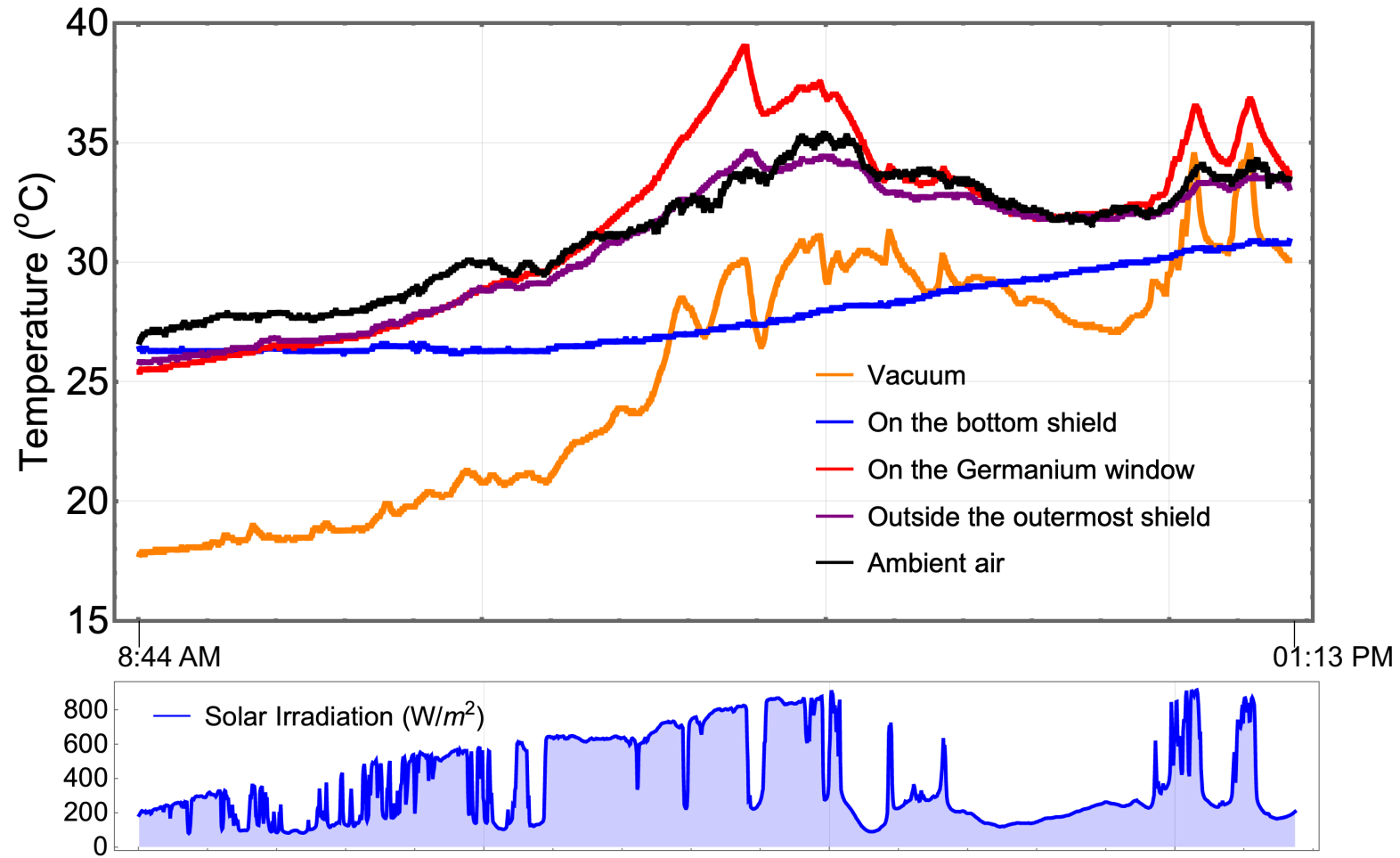

**Fig. S9**

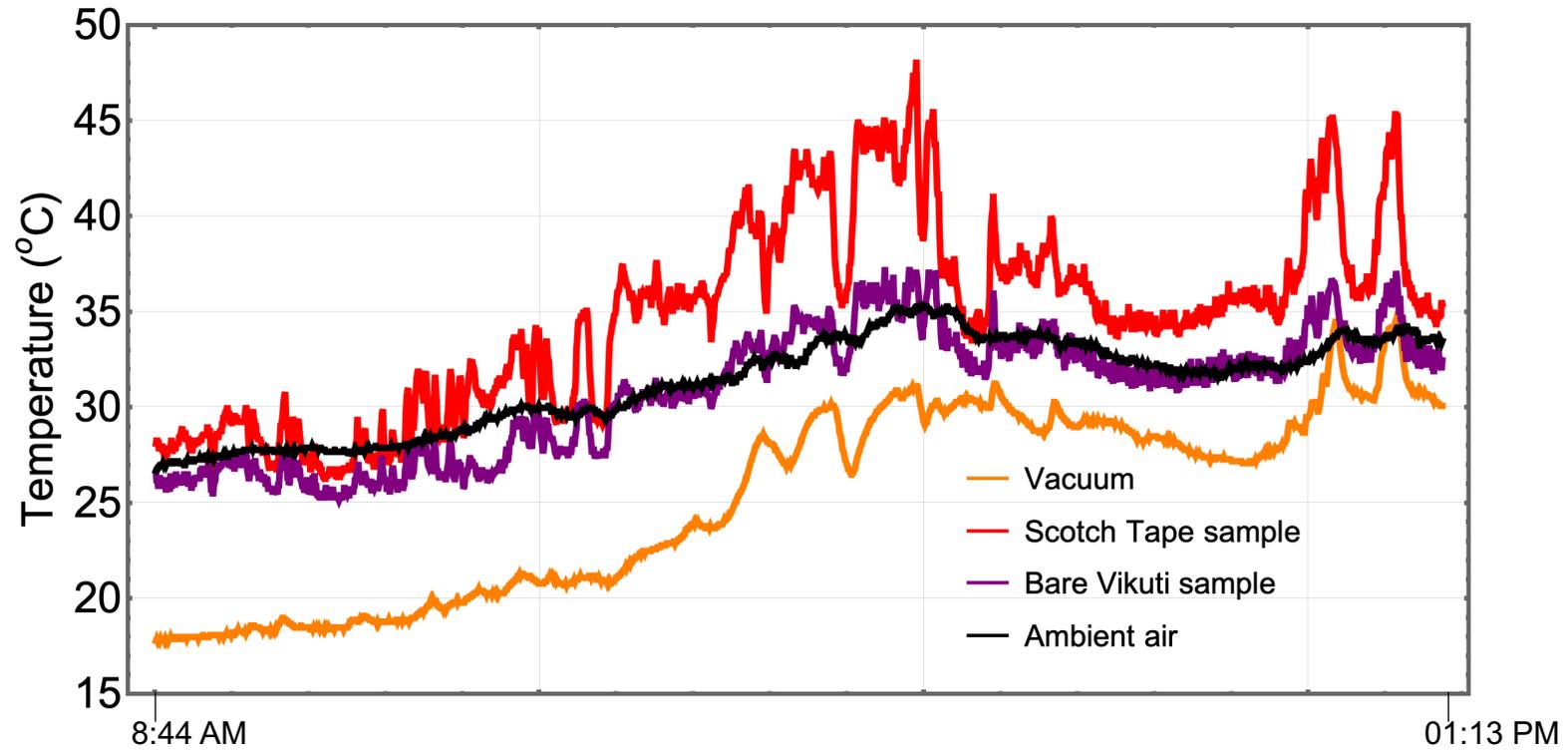

**Fig. S10**

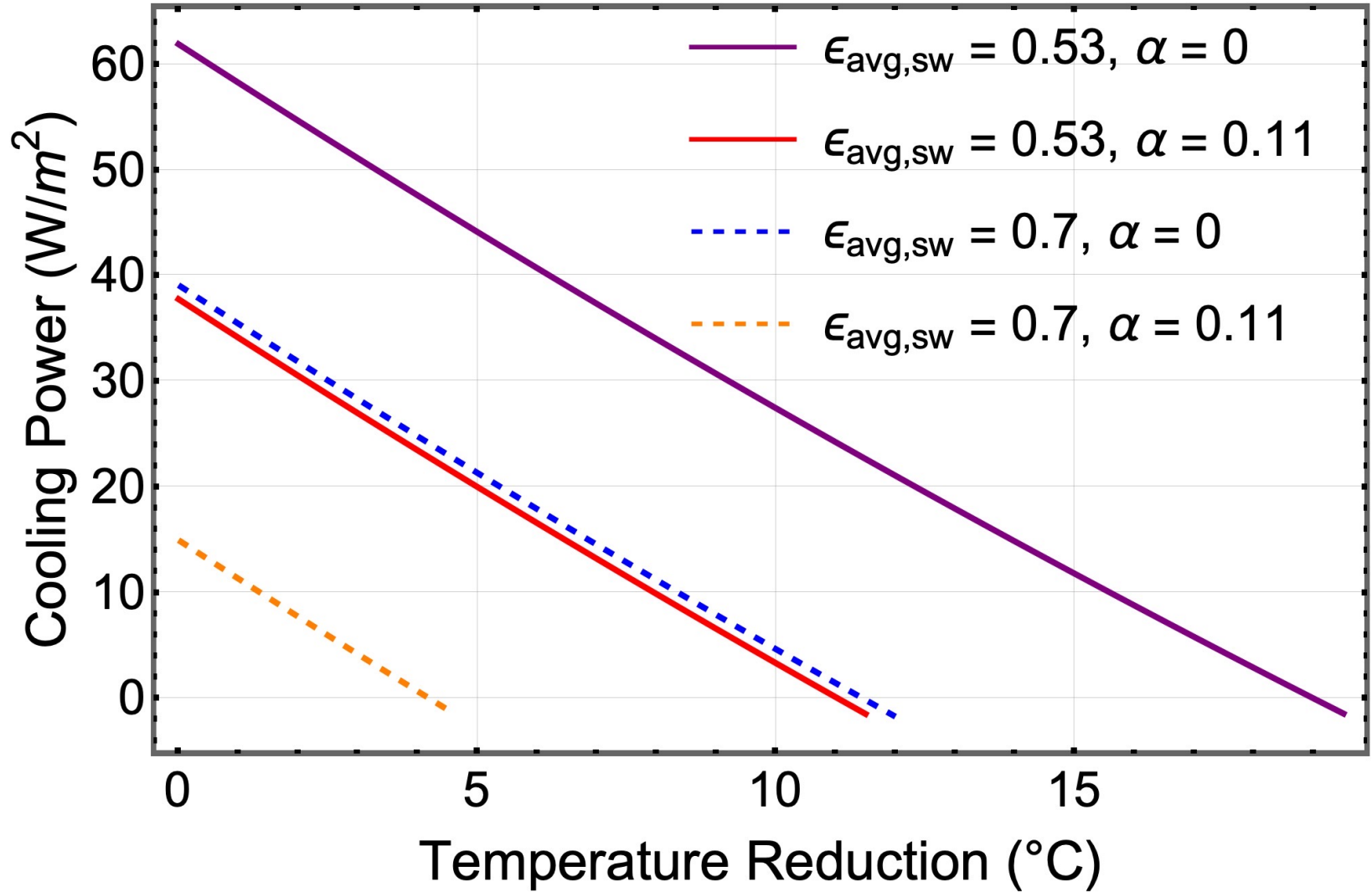